\documentclass[12pt]{article}
\usepackage{amsfonts}
\usepackage{amsmath}
\usepackage{amsbsy}
\usepackage{amssymb}
\input{epsf}

\font\grb=eurb10

\def\boldxi{\hbox{\grb\char'030}\,}

\begin{document}
\title{\bf Integrable and non-integrable structures\\ in Einstein - Maxwell equations with\\ Abelian isometry group $\mathcal{G}_2$\\[-0.5ex]}
\author{G.A.~Alekseev\footnote{e-mail: G.A.Alekseev@mi.ras.ru}\\[1ex]
\emph{Steklov Mathematical Institute of Russian Academy of Sciences}
}
\date{}
\maketitle

\begin{abstract}
The classes of electrovacuum Einstein - Maxwell fields (with a cosmological constant), which metrics admit an Abelian two-dimensional isometry group $\mathcal{G}_2$ with non-null orbits and electromagnetic fields possess the same symmetry, are considered. For the fields with such symmetries, we describe the structures of the so called "nondynamical degrees of freedom"  which presence as well as the presence of a cosmological constant change (in a strikingly similar ways) the vacuum and electrovacuum dynamical equations and destroy their well known integrable structures. The modifications of the known reduced forms of Einstein - Maxwell equations -- the Ernst equations and self-dual Kinnersley equations which take into account the presence of non-dynamical degrees of freedom are found and the subclasses of fields with different non-dynamical degrees of freedom are considered. These are: (I) vacuum metrics with cosmological constant, (II) space-time geometries in vacuum with isometry groups $\mathcal{G}_2$ which orbits do not admit the orthogonal 2-surfaces (none-orthogonally-transitive isometry groups) and (III) electrovacuum fields with more general structures of electromagnetic fields than in the known integrable cases. For each of these classes of fields, in the case of diagonal metrics, all field equations can be reduced to the only nonlinear equation of the fourth order for one real function $\alpha(x^1,x^2)$ which characterise the element of area on the orbits of the isometry group $\mathcal{G}_2$ . Simple examples of solutions for each of these classes are presented. It is pointed out that if for some two-dimensional reduction of Einstein's field equations in four or higher dimensions, the function $\alpha(x^1,x^2)$ possess a "harmonic" structure, instead of being (together with other field variables) a solution of some nonlinear equations, this can be an indication of possible complete integrability of these reduced dynamical equations for the fields with vanishing of all non-dynamical degrees of freedom.
\end{abstract}

\vfill\eject

\tableofcontents

\section{Introduction}
For deeper understanding of various mathematical structures which may arise in the studies of some classes of objects or particular models in different areas of mechanics, mathematical and theoretical physics, it may be useful to consider these classes of objects and models within more general ones because such kind of analysis can lead to some useful generalizations of the identified structures  or can show explicitly the obstacles which do not allow these structures to be extended to these more general cases.
In the present paper, this concerns the studies of various features of the structures of vacuum Einstein's and electrovacuum Einstein - Maxwell equations for the fields which are described by the known integrable reductions of these equations \footnote{The history of discovery of integrable reductions of Einstein's field equations in General Relativity -- the Einstein equations for vacuum, electrovacuum Einstein - Maxwell equations, the Einstein - Maxwell - Weyl equations and others, discussion of some related questions and  corresponding references can be found in \cite{Alekseev:2011}} and which constitute some subclass in the whole class of gravitational and elecromagnetic fields which space-times admit the Abelian isometry group $\mathcal{G}_2$ with non-null orbits and which electromagnetic fields possess the same symmetries.

\subsection{Integrability of vacuum Einstein equations with the isometry group $\mathcal{G}_2$}

In the modern theory of gravity a behaviour of strong gravitational fields and their interaction with electromagnetic fields is described respectively by vacuum Einstein equations and by electrovacuum Einstein - Maxwell equations.
Because of rather complicate nonlinear structure of these equations, the development of effective methods for analytical investigation of these equations and construction of rather rich classes of their exact solutions became possible only after a discovery of existence of integrable reductions of these equations for the fields with certain space-time symmetries. Though the integrability of these equations was conjectured much earlier  \cite{Geroch:1972} and different signs and features of this integrability (as we understand these now) could be recognized in the results of later considerations in different mathematical contexts presented by different authors \cite{Kinnersley:1977}--\cite{KCIV:1978},\cite{Maison:1978}, the actual discovery of integrability of vacuum Einstein equations was made in the paper of Belinski and Zakharov  \cite{BZ:1978} for the class of metrics which depend on time and one spatial coordinate and which possess $2\times 2$-block-diagonal structure\footnote{The similar results for stationary axisymmetric vacuum fields were published a bit later \cite{BZ:1979}.}. In \cite{BZ:1978} the authors presented a spectral problem which was equivalent to the symmetry reduced vacuum Einstein equations and which consisted of an overdetermined linear system with a spectral parameter, which compatibility conditions are satisfied for solutions of vacuum Einstein equations from the mentioned above class of metrics, and of a set of supplement conditions for the choice of solutions of this linear system providing the equivalence of the whole spectral problem to the nonlinear system of field equations.\footnote{In a bit earlier papers, Kinnersley and Chitre \cite{KCIV:1978} Maison \cite{Maison:1978} constructed some linear systems with complex parameters which compatibility conditions were satisfied by the solutions of vacuum Einstein equations. However, a formulation of a complete spectral problem which would be equivalent to the dynamical equations had not been given in these papers. Besides that, the linear system constructed in \cite{KCIV:1978} was used in another context: it was considered there as the system for a generating function for an infinite hierarchy of matrix potentials which was associated with each stationary axisymmetric solution of the field equations and which was used for a description of the action of the infinite-dimensional algebra of internal symmetries on the solutions of Einstein equations. The linear system constructed in  \cite{Maison:1978} servde as indication that vacuum Einstein equations for such fields can occur to be integrable "in the spirit of Lax".} In \cite{BZ:1978} a method was developed for generating infinite hierarchies of vacuum multi-soliton solutions, starting from arbitrary chosen already known vacuum solution of the same equations which plays the role of background for solitons.\footnote{There was a remarkable technical step also made in \cite{BZ:1978}, which is important for explicit construction and analysis of vacuum soliton solutions. This is a very simple and explicit general "determinant" expression for calculation of the conformal factor in the conformally flat part of the metric found in \cite{BZ:1978}. A compact determinant forms for all components of Belinski and Zakharov vacuum soliton solutions were found in \cite{Alekseev:1981}.}
Besides that, in \cite{BZ:1978} it was shown that for this type of vacuum metrics the other ("non-soliton") part of the space of solutions can be described in terms of the Riemann-Hilbert problem on the spectral plane and the corresponding system of linear singular integral equations.

\subsection{$\mathcal{G}_2$-symmetry of fields and integrability of Einstein - Maxwell equations}
To the time of discovery of vacuum Einstein equations it was found that the similar reductions of electrovacuum Einstein - Maxwell equations also possess the features which indicate to their complete integrability. In particular, in the framework of a group-theoretic approach, in the papers of Kinnersley and Chitre \cite{Kinnersley:1977}--\cite{KCIV:1978} it was shown the existence of infinite-dimensional algebra of internal symmetries of Einstein - Maxwell equations for stationary axisymmetric electrovacuum fields. Then, in the papers of Hauser and Ernst  \cite{HEI:1979,HEII:1979}, for more restricted class of stationary axisymmetric fields, i.e. for the fields with a regular axis of symmetry, the construction of solution generating transformations corresponding to any elements of algebra of Kinnersley and Chitre was reduced to a solution of some homogeneous Hilbert problem and equivalent system of linear singular integral equations. However, in these papers the authors had not succeded in suggesting any effective solution generating methods.\footnote{Later, a matrix system of linear singular integral equations of Hauser and Ernst was reduced to much simpler form of a scalar integral equation by Sibgatullin \cite{Sibgatullin:1984}, who restricted in these integral equations the choice of the seed solution by the Minkowski space-time  and expressed (following Hauser and Ernst \cite{HE:1980}) the Geroch group parameters in the kernel of these integral equations in terms of the values of the Ernst potentials on the axis of symmetry for generating solution. This reduced form of Hauser and Ernst integral equation was used many times by Sibgatullin with co-authors for explicit calculations of many particular examples of asymptotically flat stationary axisymmetric solutions of Einstein - Maxwell equations with the Ernst potentials which boundary values on the axis are rational functions of the Weyl coordinate $z$ along the axis. However, one can notice that all asymptotically flat solutions constructed in this way, were some particular cases of already known vacuum or electrovacuum soliton solutions generated on the Minkowski background or represented some analytcal continuations of these solitons in the spaces of their free parameters.}

Soon after the papers \cite{BZ:1978,BZ:1979} it became clear that the formulation of the inverse scattering transform and construction of vacuum soliton solutions suggested by Belinski and Zakharov do not admit a straightforward generalization to the case of interacting gravitational and electromagnetic fields. A bit later, the integrable structure of of Einstein - Maxwell equations for space-times with the isometry groups $\mathcal{G}_2$, with block-diagonal form of metric and with some specific structure of electromagnetic field components was found in the author's papers  \cite{Alekseev:1980a, Alekseev:1980b}, where a spectral problem equivalent to another (complex selfdual) form of the (symmetry reduced) Einstein - Maxwell equations was constructed. This spectral problem consists of (a) the linear system which compatibility conditions are satisfied for solutions of Einstein - Maxwell equations and (b) the supplementary conditions of existence for this linear system of the matrix integral of a special structure providing the equivalence of the complete spectral problem to electrovacuum field equations. Besides that, in the papers \cite{Alekseev:1980a, Alekseev:1980b} a method for generating of electrovacuum soliton solutions was developed. This method allows to generate the electrovacuum soliton solutions starting from arbitrary chosen known electrovacuum solution (with the same symmetry)  which plays the role of background for solitons, but technically, this method differs essentially from its vacuum analogue constructed by Belinski and Zakharov in \cite{BZ:1978,BZ:1979}. More detail description of electrovacuum solitons constructing was presented in \cite{Alekseev:1987}. On the relation between the spectral problems as well as between Belinski and Zakharov vacuum solitons and vacuum limit of electrovacuum solitons see \cite{Alekseev:2001}.

The most general approach to a description of the spaces of solutions of integrable reductions of vacuum Einstein equations and electrovacuum Einstein - Maxwell equations (called as the "monodromy transform approach") was suggested in \cite{Alekseev:1985}, where a system of linear singular integral equations
equivalent to the nonlinear dynamical equations was constructed without any supplementary restrictions on the class of vacuum or electrovacuum fields, such as regularity of the axis of symmetry or the absence of initial singularities in time-dependent solutions.
In this approach, every local solution of integrable reductions of vacuum Einstein equations or electrovacuum Einstein - Maxwell equations is characterized by  a set of coordinate independent holomorphic functions of the spectral parameter -- the monodromy data of the normalized fundamental solution of the corresponding spectral problem which enter explicitly into the kernel of the mentioned above system of linear singular integral equations. This system of integral equations admits a unique solution for any choice of the monodromy data as functions of the spectral parameter. For a special class of arbitrary rational and "analytically matched"  monodromy data functions, this system of linear integral equations admits an explicit solution  \cite{Alekseev:1987, Alekseev:1992,Alekseev-Garcia:1996}. This allows to construct infinite hierarchies of solutions of the field equations which extend essentially the classes of soliton solutions and include the solutions for some new types of field configurations.

Thus, discovery of integrable reductions of Einstein equations and Einstein - Maxwell equations and development of vacuum and electrovacuum soliton generating transformations as well as constructing of new types of field configurations solving the corresponding linear singular integral equations provide us with  effective methods for constructing the  multi-parametric families of exact solutions and finding many physically interesting examples.
In particular, these include such non-trivial field configurations as nonlinear superpositions of two charged rotating massive sources of the Kerr-Newman type \cite{Alekseev:1987}, the solution for a Schwarzschild black hole immersed into the external gravitational and electromagnetic fields of Bertotti-Robinson magnetic universe \cite{Alekseev-Garcia:1996}, equilibrium configurations of two massive charged sources of Reissner-Nordstrom type \cite{Alekseev-Belinski:2007} and others. Applications of electrovacuum soliton generating transformations was considered in \cite{Alekseev:1987}. Besides that, in the recent paper \cite{Alekseev:2016} a large class of solutions was constructed. These solutions describe a nonlinear interaction of electrovacuum soliton waves with a strong (non-soliton) pure electromagnetic waves of arbitrary amplitudes and profiles propagating in and interacting with the curvature of background space-time of homogeneously and anisotropically expanding universe which metric is described by a symmetric Kasner solution. For solutions of this class, the role of the background for solitons is played by the class of such pure electromagnetic travelling waves propagating in the symmetric Kasner background  which was found in \cite{Alekseev:2015}.

The examples of solutions given above show that the existence of integrable reductions of the field equations in the Einstein theory of gravity open the ways for development of various effctive methods for constructing of solutions. Therefore, it may be of a large interest to study the possibilities of generalization of these methods to much wider classes of fields. In this way, it may be interesting to consider those features of the structure of integrable reductions of Einstein's field equations which are retained for some more general classes of fields as well as those obstacles which destroy this integrability in more general situations. As far as for all known cases, the fields described by the integrable reductions of Einstein's field equations belong to the classes for which the space-time metric admits two-dimensional Abelian group of isometries $\mathcal{G}_2$, generated by two commuting non-null Killing vector fields, an obvious next step of our considerations is to compare the structures of the Einstein's field equations in the integrable cases with the structure of these equations for the whole classes of fields which possess the same $\mathcal{G}_2$-symmetry.

\subsection{$\mathcal{G}_2$-symmetry and non-dynamical degrees of freedom of fields}
The space-time geometries and electromagnetic field configurations, for which the Einstein - Maxwell equations were found to be  integrable, admit the Abelian two-dimensional isometry groups. However, for this integrability it is not enough to assume such symmetry. The integrable structures of these field equations were found only for some subclasses of space-times with such isometry groups.\footnote{There are different classes of space-times which admit the Abelian two-dimensional isometry groups. The difference between these classes may be in the topology and metric signature of the isometry group orbits. In particular, the orbits of Killing vector fields can be compact or not (as in the cases of axial and cylindrical symmetries or in the case of plane waves respectively) and one of two commuting Killing vector fields can be time-like and the other one -- the space-like (as for stationary axisymmetric fields) or both Killing vector fields can be space-like (as for waves and cosmological models). However, it is useful to note that this separation of solution into different classes is local: the same solution in different space-time regions can belong to different classes described above.} In the literature, further restrictions on classes of fields with $\mathcal{G}_2$-symmetry which lead eventually to integrability of the dynamical parts of this symmetry reduced field equations had arose already in different contexts and were used in different, but equivalent forms. In particular, in some cases it was assumed that two constants constructed from two commuting Killing vector fields vanish, Another (equivalent) assumption was that the orbits of the isometry group  $\mathcal{G}_2$ admit the orthogonal 2-surfaces and herefore, in appropriate coordinates metric takes $2\times 2$-block-diagonal form  \cite{Lewis:1932,Papapetrou:1966, Carter:1972}.

Despite the difficulties with integration of field equations for more general classes of fields than those described by integrable reductions of Einstein's field equations, these equations also have been considered in the literature. In particular, a set of examples of vacuum and electrovacuum solutions with cosmological constant can be found (with the corresponding references) in the books \cite{SKMHH:2009,Griffiths-Podolsky:2009}. Rather cumbersome formalizm for the analysis of Einstein equations for space-times which admit the two-dimensional non-orthogonally-transitive Abelian isometry group was suggested in \cite{Gaffet:1990}. A more general (than in the integrable cases) structure of Einstein - Maxwell equations was considered in  \cite{Gourgoulhon-Bonazzola:1993} where it was assumed that the energy-momentum tensor of matter sources does not satisfy the so called "circularity" condition (see \cite{Carter:1972}) which is the necessary condition for stationary axisymmetric metric to possess a block-diagonal form. It was noticed there also that this class of fields certainly represent a physical interest because the solutions of this type should arise in the case of stationary axisymmetric field configurations with toroidal magnetic fields.

In the present paper, the structures of vacuum Einstein equations and electrovacuum Einstein - Maxwell equations are considered for the whole classes of space-times which admit two-dimensional Abelian isometry groups  $\mathcal{G}_2$. For these classes of fields we show, that besides the dynamical variables these electromagnetic and gravitation fields possess some "non-dynamical degrees of freedom" which we use further in order to derive the necessary conditions for the mentioned above integrability in more physical terms. The term "non-dynamical degrees of freedom" means here those metric and electromagnetic field components for which we do not have the dynamical equations but instead of these, the part of Einstein equations, which represent the constraints, admit the explicit solutions which allow to express these field components in terms of the other (dynamical) variables and a set of arbitrary constant parameters which arise from the constraint equations as the constants of integration. These non-dynamical degrees of freedom  can possess an important physical interpretation. For example, for stationary axisymmetric fields, two electromagnetic non-dynamical degrees of freedom describe the vortex parts of electric and magnetic fields determined by non-vanishing azimuthal components which are absent in the case of solutions of integrable reductions of Einstein - Maxwell equations.

As we shall see below,  the arbitrary constants, which characterize the non-dynamical degrees of freedom of gravitational and electromagnetic fields with $\mathcal{G}_2$-symmetries, can enter also into the symmetry reduced dynamical parts of vacuum Einstein equations and electrovacuum Einstein - Maxwell equations as free parameters and change crucially the structure of these dynamical equations. It is interesting to note that the form in which these parameters enter the dynamical part of the field equations is very similar to the case of presence of cosmological constant. That is why in this paper, we include the cosmological constant in our considerations as one more non-dynamical degree of freedom of the gravitational field. The presence of these arbitrary constants, similarly to the case of presence of the cosmological constant, destroy the well known today integrable structures of vacuum Einstein equations and electrovacuum Einstein - Maxwell equations, however, it is necessary to appreciate that the question of integrability (probably, in some other form)  of these more general equations remains to be opened.

Nonetheless, an important property of arbitrary cosntants which determine the non-dynamical degrees of freedom of gravitational and electromagnetic fields in a general classes of vacuum and electrovacuum spacetimes with $\mathcal{G}_2$-symmetries is that for some particular values of these constants the corresponding field components, identified with the non-dynamical degrees of freedom, become pure gauge and threfore, these components can vanish for appropriate choice of coordinates and gauges. In this case, the dynamical equations reduce to those which were found earlier to be integrable. Thus, we show that the necessary and sufficient condition providing the integrability of $\mathcal{G}_2$-symmetry reduced vacuum Einstein equations and electrovacuum Einstein - Maxwell equations can be presented in an equivalent but more physical form as the condition of vanishing of all non-dynamical degrees of freedom of gravitational and electromagnetic fields.

Further, we consider different subclasses of fields with non-dynamical degrees of freedom: (I) vacuum metrics with non-vanishing cosmological constant, (II) space-time geometries in which the orbits of the isometry group $\mathcal{G}_2$ do not admit the ortogonal 2-surfaces, i.e. the isometry group $\mathcal{G}_2$ is not orthogonally-transitive and (III) electrovacuum fields жиҐа more general structures of electromagnetic fields than in the known integrable cases. For each of these subclasses, in the case of diagonal metrics all field equations can be reduced to one nonlinear equation for one real function $\alpha(x^1,x^2)$, which possesses a geometrical interpretation as describing the element of the area on the orbits of the isometry group $\mathcal{G}_2$. The simple examples of solutions are geven here for each of these subclasses of fields with non-vanishing non-dynamical degrees of freedom.

In the Concluding remarks, we summarized the results of this paper concerning the "status" of fields described by integrable reductions of vacuum Einstein equations and electrovacuum Einstein - Maxwell equations within the corresponding general classes of vacuum or electrovacuum fields with a cosmological constant in space-times which metrics admit two-dimensional Abelian isometry group and electromagnetic fields possess the same symmetry.

\section{Electrovacuum space-times with two-dimensional\\ Abelian isometry group $\mathcal{G}_2$}
In General Relativity, the dynamics of gravitational and electromagnetic fields outside their sources is determined by electrovacuum Einstein - Maxwell equations ($\gamma=c=1$):
\begin{equation}\label{EMEquations}
\begin{array}{l}R_{ik}-\dfrac 12 R g_{ik}+\Lambda g_{ik}=8\pi T^{\scriptscriptstyle{\textsc{(M)}}}_{ik},\quad
T^{\scriptscriptstyle{\textsc{(M)}}}_{ik}=\dfrac 1{4\pi}(F_{il} F_k{}^l-\frac 14 F_{lm}F^{lm} g_{ik})\\[2ex]
\nabla_k F_i{}^k=0,\hskip3ex  \nabla_{[i} F_{jk]}=0,
\end{array}
\end{equation}
where $i,j,k,\ldots=0,1,2,3$, metric signature is $(-+++)$ and $\Lambda$
is a cosmological constant. Most of the known today exact solutions of these equations possess the two-dimensional symmetries. We consider here the whole class of solutions of (\ref{EMEquations}) which admit a two-dimensional Abelian isometry group which action is generated by two commuting non-null Killing vector fields, denoted further as the isometry group $\mathcal{G}_2$.

\subsection{Basic notations and definitions}

\paragraph{\rm\textit{\underline{Orbit space and the choice of coordinates.}}} If the space-time admits two-dimensional Abelian isometry group $\mathcal{G}_2$, its action foliates this spacetime by two-dimensional non-null surfaces -- the orbits of this group. A factor of this space-time manifold with respect to the action of $\mathcal{G}_2$ is a two-dimensional manifold called as the orbit space of this isometry group. In  such space-times, the coordinate systems can be chosen so that the coordinates $x^1$ and $x^2$ are the coordinates on the orbit space and another  pair of coordinates $x^3$ and $x^4$ are the  coordinates on the orbits. It is convenient to choose the coordinates $x^3$ and $x^4$ as natural parameters on the lines of two commuting Killing vector fields which generate the action of $\mathcal{G}_2$ and are tangent to its orbits. In these coordinates the components of a chosen pair of Killing vector fields  take the forms
\begin{equation}\label{Killings}
\boldxi_{(3)}=\frac \partial{\partial x^3},\quad \boldxi_{(4)}=\frac \partial{\partial x^4}\qquad\text{or}\qquad \xi^k{}_{(a)}=\delta^k{}_a
\end{equation}
where the index in parenthesis $a=3,4$ numerates the Killing vector fields and corresponds to the coordinate chosen as the parameter along line of this Killing vector field. For such choice of coordinates, metric and electromagnetic field components depend only on the coordinates $x^\mu=\{x^1,x^2\}$ and these are independent of $x^a=\{x^3,x^4\}$.

\paragraph{\rm\textit{\underline{\textit{Metric and electromagnetic field components.}}}} In the space-time with two commuting isometries (\ref{Killings}), in the coordinates described above, the components of metric and the Maxwell tensor of electromagnetic field  can be presented in a general form
\begin{equation}\label{gFComponents}
\begin{array}{l}
ds^2=g_{\mu\nu} dx^\mu dx^\nu +g_{ab} (dx^a+\omega^a{}_\mu dx^\mu)( dx^b+\omega^b{}_\nu dx^\nu),\\[2ex]
\left.F_{ik}=\begin{pmatrix}
F_{\mu\nu}&F_{\mu b}\\[1ex]
F_{a \nu}& F_{ab}
\end{pmatrix},\hskip10ex\right\Vert\qquad
\begin{array}{l}
\mu,\nu,\ldots=1,2\\[0.5ex]
a,b,\ldots=3,4
{}
\end{array}
\end{array}
\end{equation}
where $g_{\mu\nu}$, $g_{ab}$, $\omega^a{}_\mu$, $F_{\mu\nu}$, $F_{\mu a}=-F_{a\mu}$ and $F_{ab}$ depend on $x^\mu=\{x^1,x^2\}$ only.
It is important that we do not preset in advance if the time coordinate is among $(x^1,x^2)$ or it is among $(x^3,x^4)$ and therefore, we consider in a unified manner both possible classes of fields -- the stationary fields with one spatial isometry and time-dependent fields with two spatial isometries.

\paragraph{\rm\textit{\underline{Conformal coordinates on the orbit space of $\mathcal{G}_2$.}}} In (\ref{gFComponents}), the components $g_{ab}$ determine the metric on the orbits of the isometry group $\mathcal{G}_2$ and the components $g_{\mu\nu}$ -- the metric on the orbit space. It is well known that on a two-dimensional surface the local coordinates can be chosen so that its metric takes a conformally flat form. We use this convenient choice of local coordinates on the orbit space $x^\mu=\{x^1,x^2\}$ for which
\begin{equation}\label{confact}
g_{\mu\nu}=f\,\eta_{\mu\nu},\qquad \eta_{\mu\nu}=\begin{pmatrix}\epsilon_1&0\\
0&\epsilon_2\end{pmatrix},\quad \begin{array}{l} \epsilon_1=\pm 1,\\
\epsilon_2=\pm 1,\end{array}\qquad \epsilon\equiv -\epsilon_1\epsilon_2
\end{equation}
where, by definition, $f > 0$, and we use the sign symbols $\epsilon_1$ and $\epsilon_2$ to consider in a unified manner all possible cases of the signature of metric on the orbits. Namely, for stationary fields for which the time coordinate is among the coordinates $x^a=\{x^3,x^4\}$ and both coordinates $x^\mu$ are space-like, the values of the  symbols $\epsilon_1$ and $\epsilon_2$ coincide, but for other cases, in which the time coordinate is among the coordinates $x^\mu=\{x^1,x^2\}$, the symbols $\epsilon_1$ and $\epsilon_2$ take the opposite values. Therefore, for the sign symbol $\epsilon$ introduced in (\ref{confact}) and for its "square root" $j$  which will be used further we have
\[\epsilon=\left\{\begin{array}{ll}
\phantom{-}1&\text{-- for time-depemdemt fields,}\\
-1&\text{-- for stationary fields,}
\end{array}\right. \qquad
j=\left\{\begin{array}{lcll}
1,&&\text{for}& \epsilon=1,\\
i,&&\text{for}& \epsilon=-1.
\end{array}\right.
\]
We call also the case $\epsilon=1$ as the hyperbolic case and $\epsilon=-1$ as the elliptic case.
Assuming that the 4-dimensional spacetime metric (\ref{gFComponents}) possess a maximal positive Lorenz signature  $(-+++)$, we have $\epsilon_1=-1$, $\epsilon_2=1$ for $\epsilon=1$ and $\epsilon_1=\epsilon_2=1$ for $\epsilon=-1$.
For Levi-Civita symbol determined on the orbit space $x^\mu=\{x^1,x^2\}$ we have:
\[\varepsilon^{\mu\nu}=\begin{pmatrix}0&1\\ -1&0\end{pmatrix},\qquad
\varepsilon_\mu{}^\nu=\eta_{\mu\gamma}\varepsilon^{\gamma\nu}=
\begin{pmatrix}0&\epsilon_1\\ -\epsilon_2&0\end{pmatrix},\qquad
\varepsilon_{\mu\nu}=\varepsilon_\nu{}^\gamma\eta_{\gamma\nu}=-\epsilon\begin{pmatrix}0&1\\ -1&0\end{pmatrix}.
\]
The conditions (\ref{confact}) determine the choice of coordinates $x^\mu$ up to arbitrary transformations of the form $(x^1+j x^2)\to A(x^1+j x^2)$ and $(x^1-j x^2)\to B(x^1-j x^2)$ which transform the conformal factor $f$ defined in (\ref{confact}) as $f\to A^\prime B^\prime f$. This choice of coordinates can be specified more, using for these coordinates the functions determined by space-time geometry itself.

\paragraph{\rm\textit{\underline{Metric on the orbits of $\mathcal{G}_2$.}}} The $2\times 2$-matrix of the components $g_{ab}$  of the metric (\ref{gFComponents}) describes the metric on the orbits of the isometry group $\mathcal{G}_2$. In the symmetry reduced Einstein - Maxwell equations, these components will play the role of dynamical variables for gravitational field. However, it is more convenient to use instead of these variables, a function $\alpha > 0$ -- the element of the area on the orbits defined as
\begin{equation}\label{Alpha}
  \det\Vert g_{ab}\Vert\equiv\epsilon\alpha^2
\end{equation}
and a $2\times 2$-matrix $\Vert h\Vert$ which components with lower indices coincide with those for $\Vert g\Vert$: $h_{ab}\equiv g_{ab}$, but  following Kinnersley \cite{Kinnersley:1977}, its indices will be rised and lowered like the two-component spinor indeces.Then, the components of $\Vert h\Vert$ with lower, mixed and upper indices will possess the expressions
\begin{equation}\label{hab}
\begin{array}{lcl}
h_{ab}\equiv g_{ab},&& h_a{}^b=-h_{ac}\epsilon^{cb},\\[1ex]
h^{ab}=\epsilon^{ac} h_c{}^b, && h_{ab}=h_a{}^c \epsilon_{cb},
\end{array}\quad\text{where}\quad \epsilon^{ab}=\epsilon_{ab}=\begin{pmatrix}
0&1\\ -1&0\end{pmatrix}
\end{equation}
It is worth to mention here also some useful relations:
\begin{equation}\label{hh}
\det\Vert h_a{}^b\Vert=-\epsilon\alpha^2,\qquad h_c{}^c=0,\qquad h_a{}^c h_c{}^b=-\epsilon\alpha^2 \delta_a{}^b,\qquad g^{ab}=\epsilon\alpha^{-2} h^{ab}.
\end{equation}

\paragraph{\rm\textit{\underline{Components of the Ricci tensor for metrics (\ref{gFComponents}).}}}
In our notations, the components of the Ricci tensor for metrics (\ref{gFComponents}) possess the expressions
\begin{equation}\label{Ricci}
\begin{array}{l}
R_\mu{}^\nu=\dfrac 1f\bigl[
-\dfrac 12 \partial_\delta\bigl(\dfrac{f^\delta}f\bigr)\eta_{\mu\gamma}+\dfrac {f_\gamma\alpha_\mu+f_\mu\alpha_\gamma-\alpha_\delta f^\delta\eta_{\mu\gamma}}{2 f \alpha}-\dfrac{\alpha_{\mu\gamma}}{\alpha}- \dfrac{\epsilon}{4\alpha^2}\partial_\mu h_c{}^d\partial_\gamma h_d{}^c\bigr]\eta^{\gamma\nu}\\[3ex]
\phantom{R^\nu{}_\mu=}+\dfrac\epsilon{2 f^2}\left(\mathcal{T}^c h_{cd}\mathcal{T}^d\right)\delta_\mu{}^\nu+\omega^c{}_\mu R_c{}^\nu{},\\[2ex]
R_a{}^\mu= -\dfrac{\epsilon}{2 f\alpha}\varepsilon^{\mu\gamma}\partial_\gamma\left(\dfrac\alpha f h_{ac}\mathcal{T}^c\right),\qquad\text{where}\qquad
\mathcal{T}^a\equiv\varepsilon^{\mu\nu}\partial_\mu\omega^a{}_\nu,
\\[3ex]
R_a{}^b=-\dfrac \epsilon{2 f\alpha} \eta^{\gamma\delta}\partial_\gamma
\bigl[\alpha\partial_\delta g_{ac} g^{cb}\bigr]-\dfrac{\epsilon}{2 f^2}g_{ac}\mathcal{T}^c\mathcal{T}^b- R_a{}^\gamma\omega^b{}_\gamma,
\end{array}
\end{equation}
where $f_\mu=\partial_\mu f$, $\alpha_\mu=\partial_\mu \alpha$, $\alpha_{\mu\nu}=\partial_\mu\partial_\nu\alpha$ and $f^\gamma=\eta^{\gamma\delta} f_\delta$; the functions $f > 0$ and $\alpha > 0$ were defined respectively in (\ref{confact}) and (\ref{Alpha}).

\paragraph{\rm\textit{\underline{Self-dual Maxwell tensor and its complex vector potential.}}}The Maxwell tensor and its dual conjugate can be combined in the self-dual Maxwell tensor
\begin{equation}\label{selfdualMaxwell}
{\overset + F}{}_{ik}=F_{ik}+i {\overset \ast F}{}_{ik},\qquad
 {\overset \ast F}{}^{ik}=\dfrac 12\, \varepsilon^{iklm} F_{lm},\qquad
{\overset + F}{}^{ik}=\dfrac i 2 \varepsilon^{iklm} {\overset + F}{}_{lm},
\end{equation}
where $\varepsilon^{iklm}=e^{iklm}/\sqrt{-g}$ is the four-dimensional Levi-Civita tensor and $e^{iklm}$ is a four-dimensional antisymmetric Levi-Civita symbol such that $e^{0123}=1$. For this tensor, a pair of Maxwell equations in  the second line of (\ref{EMEquations}) can be combined in a complex equation which is equivalent to the existence of a complex vector potential
\begin{equation}\label{dualMaxwell}
 \varepsilon^{iklm}\nabla_k {\overset + F}{}_{lm}=0,\qquad\Rightarrow\qquad {\overset + F}{}_{ik}=\partial_i \Phi_k-\partial_k \Phi_i.
\end{equation}

For spacetimes which admit the isometry group $\mathcal{G}_2$, provided the electromagnetic fields share this symmetry, the Lie drivatives of the Maxwell tensor along the Killing vector fields $\boldxi_{(a)}$ vanish. This means that in the coordinates (\ref{gFComponents}) the components of self-dual Maxwell tensor are independent of the coordinates $x^a$, but in general, this should not be necessary true for the components of its vector potential $\Phi_i$. In the coordinates (\ref{gFComponents}), the components of $\Phi_i$ split in two parts $\Phi_i=\{\Phi_\mu,\,\Phi_a\}$ so that
\begin{equation}\label{PhimuPhia}
{\overset + F}{}_{\mu\nu}=\partial_{\mu} \Phi_{\nu}-\partial_{\nu} \Phi_{\mu},\qquad
{\overset + F}{}_{\mu b}=\partial_{\mu} \Phi_{b}-\partial_{b} \Phi_{\mu},\qquad {\overset + F}{}_{bc}=\partial_{b} \Phi_{c}-\partial_{c} \Phi_{b}.
\end{equation}
For later convenience, we present here also the general expressions for the components of self-dual Maxwell tensor with  upper indices
\begin{equation}\label{Fplusuu}
\begin{array}{l}
{\overset + F}{}^{\mu\nu}=\dfrac 1{f^2}\eta^{\mu\gamma}\left(
{\overset + F}{}_{\gamma\delta}+\omega^c{}_\gamma {\overset + F}{}_{\delta c}-{\overset + F}{}_{\gamma c} \omega^c{}_\delta+\omega^c{}_\gamma {\overset + F}{}_{cd}\omega^d{}_\delta\right)\eta^{\delta\nu}\\
{\overset + F}{}^{\mu a}=-{\overset + F}{}^{\mu\gamma}\omega^a{}_\gamma+
\dfrac 1 {\epsilon \alpha^2 f}\eta^{\mu\gamma}({\overset + F}{}_{\gamma d}-\omega^c{}_\gamma {\overset + F}{}_{cd})h^{da}\\
{\overset + F}{}^{ab}=\dfrac 1{\alpha^4} h^{ac}{\overset + F}{}_{cd} h^{db}- \dfrac1{\epsilon\alpha^2 f}h^{ac}({\overset + F}{}_{c\gamma}-{\overset + F}{}_{cd}\omega^d{}_\gamma)\eta^{\gamma\delta}\omega^b{}_\delta\\
\phantom{{\overset + F}{}^{ab}=} -\dfrac 1{\epsilon\alpha^2 f}\omega^a{}_\gamma\eta^{\gamma\delta}({\overset + F}{}_{\delta c}-\omega^d{}_\delta {\overset + F}{}_{dc}) h^{cb}+\omega^a{}_\gamma {\overset + F}{}^{\gamma\delta}\omega^b{}_\delta.
\end{array}
\end{equation}

To write the equations (\ref{EMEquations}) for spacetimes with metric and electrpmagnetic fields (\ref{gFComponents}), we need now the expressions for the components of the energy-momentum tensor.

\paragraph{\rm\textit{\underline{The energy-momentum tensor for electromagnetic fields.}}} It is easy to check that in general the energy-momentum tensor for electromagnetic field (\ref{EMEquations}) can be  expressed in terms of self-dual Maxwell tensor and its complex conjugate:
\begin{equation}\label{Tik}
T^{\scriptscriptstyle{\textsc{(M)}}}{}_i{}^k=-\dfrac1{8\pi}\overline{{\overset + F}}{}_{il}{\overset + F}{}^{lk}
\end{equation}
In the coordinates (\ref{gFComponents}) the components of this tensor possess the expressions
\begin{equation}\label{TmunuTmuaTab}
\begin{array}{l}
T^{{\scriptscriptstyle{\textsc{(M)}}}}{}_\mu{}^\nu = \dfrac{i}{8\pi f\alpha}\bigl[
e_o(\varepsilon^{\gamma\delta}\partial_\gamma\overline{\Phi}{}_\delta) \delta_\mu{}^\nu{}- \varepsilon^{\nu\gamma}\partial_\gamma\Phi^c\partial_\mu\overline{\Phi}_c\bigr],\\[2ex]
T^{{\scriptscriptstyle{\textsc{(M)}}}}{}_a{}^\mu=\dfrac i{8\pi f\alpha}\varepsilon^{\mu\gamma}\partial_\gamma\bigl(
\overline{e}{}_o \Phi_a-e_o \overline{\Phi}_a\bigr),\\[2ex]
T^{{\scriptscriptstyle{\textsc{(M)}}}}{}_a{}^b=\dfrac{i}{8\pi f \alpha}\varepsilon^{\gamma\delta}\bigl[\partial_\gamma\Phi_a\partial_\delta \overline{\Phi}{}^b- e_o(\partial_\gamma\overline{\Phi}{}_\delta)\delta_a{}^b\bigr].
\end{array}
\end{equation}
Now, the Einstein - Maxwell field equations (\ref{EMEquations}) can be written, using the above notations and the expressions (\ref{Ricci}) and (\ref{TmunuTmuaTab}).

\subsection{$\mathcal{G}_2$-symmetry reduced Einstein - Maxwell field equations}
In this section we consider a complete set of Einstein - Maxwell equations for electrovaccum fields (with a cosmological constant $\Lambda$) which admit an Abelian isometry group $\mathcal{G}_2$. We describe the structure of these equations and determine their constraint and dynamical parts. The general solution of constraint equations presented below in this section allows to determine the non-dynamical degrees of freedom of gravitational and electromagnetic fields and find the general form of $\mathcal{G}_2$-symmetry reduced dynamical equations. Finally, we show in this section that for some special values of integration constants and $\Lambda=0$, the non-dynamical degrees of freedom become pure gauge and vanishing for appropriate choice of gauges and coordinates. In this case, it occurs that the physical condition of vanishing of non-dynamical degrees of freedom of gravitational and electromagnetic fields is equivalent to pure geometrical condition of $2\times 2$-block-diagonal form of metric which reduce the general classes of electrovacuum fields with $\mathcal{G}_2$-symmetries to the corresponding subclasses which dynamical equations are completely integrable.

\subsubsection{Non-dynamical degrees of freedom of electromagnetic field} It is easy to see that the independence of the components of self-dual Maxwell tensor on the coordinates $x^a$ implies a specific structure of some of the components of this tensor:
\begin{equation}\label{eo}
{\overset + F}_{ab}= e_o \epsilon_{ab},\quad e_o=const,
\end{equation}
where $e_o$ is an arbitrary complex constant. This means that the corresponding components of complex vector potential $\Phi_i$ are the linear functions of the coordinates $x^a$. Up to arbitrary gauge transformation, these components possess the structure
\begin{equation}\label{PhimuPhia1}
\Phi_\mu=\Phi_\mu(x^\gamma),\qquad
\Phi_a=\dfrac 12 e_o\epsilon_{ca} x^c+\phi{}_a(x^{\gamma}).
\end{equation}
In this class of fields, the real and imaginary parts of arbirary complex constant $e_0$ can be considered as determining two non-dynamical degrees of freedom of electromagnetic field. The corresponding components of electromagnetic potential representing these degrees of freedom can be calculated from the constraint part of self-dual Maxwell equations.

\subsubsection{Self-dual form of symmetry-reduced Maxwell equations}
In terms of a complex vector potential the Maxwell equations (\ref{dualMaxwell}) are equivalent to the first order duality equations
\begin{equation}\label{sdMaxwell}
{\overset + F}{}^{ik}=\dfrac i 2 \varepsilon^{iklm} {\overset + F}{}_{lm},\qquad {\overset + F}{}_{ik}=\partial_i \Phi_k-\partial_k \Phi_i.
\end{equation}
In the coordinates (\ref{gFComponents}), using the expressions (\ref{Fplusuu}) and (\ref{PhimuPhia1}), we can split the duality equations (\ref{sdMaxwell}) in two parts. One of them is the so called constraint equation
\begin{equation}\label{Phimu}
\varepsilon^{\mu\nu}\partial_\mu\Phi_\nu= \varepsilon^{\mu\nu}\partial_\mu\Phi_c\,\omega^c{}_\nu-e_o\bigl[\dfrac12 \varepsilon^{\mu\nu}\epsilon_{cd}\omega^c{}_\mu\omega^d{}_\nu+i \epsilon \alpha^{-1} f\bigr],
\end{equation}
One can see here that the components $\Phi_\mu$ of a complex electromagnetic potential represent the non-dynamical degrees of freedom of electromagnetic field because these components are determined (up to arbitrary gauge transformation $\Phi_\mu\to\Phi_\mu+\partial_\mu\chi$) by the equation (\ref{Phimu}), provided the solution for $\Phi_a$, $\alpha$, $\omega^a{}_\mu$ and $f$ is known. Another part of (\ref{sdMaxwell}) in the metrics (\ref{gFComponents}) reduce to a system of duality  equations (similarly to the Kinnersley equations \cite{Kinnersley:1977})
\begin{equation}\label{sdMaxwelleqs}
\partial_\mu \Phi_a-e_o \epsilon_{ca}\omega^c{}_\mu=i\alpha^{-1}\varepsilon_\mu{}^\gamma h_a{}^c (\partial_\gamma \Phi_c-e_o \epsilon_{dc}\omega^d{}_\gamma),
\end{equation}
where the components $\Phi_a$ of a complex potential play the role of dynamical variables.

\subsubsection{Non-dynamical degrees of freedom of gravitational field}
The elecrovacuum Einstein - Maxwell field equations (\ref{EMEquations}) for spacetimes with $\mathcal{G}_2$-symmetries can be written
using the expressions (\ref{Ricci}), (\ref{Tik}), (\ref{TmunuTmuaTab}).
In this form, these equations consist of three subsystems coupled to each other, but possessing different structures.

\paragraph{\rm\textit{\underline{Three subsystems of Einstein-Maxwell equations.}}}
For electrovacuum fields the trace $T^{{\scriptscriptstyle{\textsc{(M)}}}k}{}_k$ of the energy-momentum tensor vanishes and therefore, the curvature scalar $R=4\Lambda$ and electrovacuum Einstein - Maxwell equations can be written in the form
\begin{equation}\label{RRR}
R_\mu{}^\nu=8\pi T^{{\scriptscriptstyle{\textsc{(M)}}}}{}_\mu{}^\nu+\Lambda \delta_\mu{}^\nu,\qquad R_a{}^\nu=8\pi T^{{\scriptscriptstyle{\textsc{(M)}}}}{}_a{}^\nu,\qquad
R_a{}^b=8\pi T^{{\scriptscriptstyle{\textsc{(M)}}}}{}_a{}^b+\Lambda \delta_a{}^b,
\end{equation}
In general, for spacetimes with the isometry group $\mathcal{G}_2$, the three subsystems of equations (\ref{RRR}) are coupled to each other and possess rather complicate structure. However, these structures can be simplified when one will observe that the second subsystem in (\ref{RRR}) represents a constraint equations which can be solved explicitly and its solution allows to identify some "non-dynamical" degrees of freedom for gravitational field.

\paragraph{\rm\textit{\underline{Constraint equations for $\omega^a{}_\mu$.}}}
The second subsystem in (\ref{RRR}), in accordance with the\\ expressions (\ref{Ricci}) and (\ref{TmunuTmuaTab}), takes the form of constraint equations
\begin{equation}\label{ka}
\partial_\mu\Bigl[\dfrac\alpha f h_{ac}\mathcal{T}^c+
2 i\epsilon (\overline{e}{}_o \Phi_a-e_o \overline{\Phi}_a)\Bigr]=0
\end{equation}
which can be solved explicitly and the general solution of these equations includes  arbitrary real two-dimensional vector constant of integration:
\begin{equation}\label{TTT}
\dfrac\alpha f h_{ac}\mathcal{T}^c=
-2 i\epsilon(\overline{e}{}_o \Phi_a-e_o \overline{\Phi}_a)+\ell_a,\qquad\text{where}\qquad  \ell_a=const.
\end{equation}
This relation allows to exclude $\mathcal{T}^a$ from the other equations (\ref{RRR}) and besides that, using the definition of $\mathcal{T}^a$ given in (\ref{Ricci}) we obtain the equation for $\omega^a{}_\mu$ of the form
\begin{equation}\label{omegas}
\varepsilon^{\mu\nu}\partial_\mu\omega^a{}_\nu=\epsilon \alpha^{-3} f h^{ac}
L_c,\qquad L_a=\ell_a-2 i\epsilon(\overline{e}{}_o \Phi_a-e_o \overline{\Phi}_a)
\end{equation}
which allows to calculate the metric functions $\omega^a{}_\mu$ (up to an arbitrary  gauge transformation $\omega^a{}_\mu\to\omega^a{}_\mu+\partial_\mu \omega^a$) in terms of  $h_{ab}$, $\Phi_a$, $f$ and the constant vector $\ell_a$. Thus, we can consider the metric functions $\omega^a{}_\mu$, together with the components $\Phi_\mu$ of complex electromagnetic potential, as the non-dynamical degrees of fredom for gravitational and electromagnetic fields  determined (up to the gauge transformations) by the constraint equations (\ref{omegas}) and (\ref{Phimu}) respectively with the arbitrarily chosen constants $\ell_a$ and $e_o$.

\paragraph{\rm\textit{\underline{Constraint equations for the conformal factor $f$}}}
With the constraint equations (\ref{Phimu}) and (\ref{omegas}), the first  subsystem in (\ref{RRR}) reduces to
\begin{equation}\label{fconstraint}
\dfrac {f_\mu\alpha_\nu+f_\nu\alpha_\mu}{f}=\mathcal{F}_{\mu\nu}+ X\eta_{\mu\nu}\\[1ex]
\end{equation}
where the components of $\mathcal{F}_{\mu\nu}$ are independent of $f$Ў and the function $X$ includes the second derivatives of $f$. These functions  possess rather long expressions:
\begin{equation}\label{FXstructure}
\begin{array}{l}
\mathcal{F}_{\mu\nu}\equiv 2\alpha_{\mu\nu}+ \dfrac{\epsilon}{2\alpha}\partial_\mu h_c{}^d\partial_\nu h_d{}^c+ i\varepsilon_\nu{}^\gamma\Bigl[M_\gamma{}^c\epsilon_{cd} \overline{M}_\mu{}^d-
\overline{M}_\gamma{}^c\epsilon_{cd} M_\mu{}^d\Bigr]
,\\[2ex]
X\equiv \partial_\delta\bigl(\dfrac{\alpha f^\delta}f\bigr)+2 f\alpha\Lambda-2\epsilon e_o \overline{e}_o\dfrac {f}{\alpha}
-\dfrac{f}{\alpha^3}(L_c h^{cd} L_d).
\end{array}
\end{equation}
where $\mathcal{F}_{\mu\nu}$ is symmetric, $X$ and $\mathcal{F}_{\mu\nu}$ are real and the components й  $M$ are defined as
\begin{equation}\label{Mamu}
M_\mu{}^a\equiv \partial_\mu\Phi^a -e_o \omega^a{}_\mu,\qquad M_{\mu a}=M_\mu{}^c \epsilon_{c a}.
\end{equation}
Instead of the equations (\ref{fconstraint}), it is more convenient to consider an equivalent system which consists of the non-diagonal component $(\ldots)_{12}$, the combination $(\ldots)_{11}+\epsilon (\ldots)_{22}$ and a contraction of (\ref{fconstraint}) with $\eta^{\mu\nu}$, i.e. the equation $(\ldots)_{11}-\epsilon (\ldots)_{22}$. It is easy to see that the first two of these equations do not include the second derivatives of $f$ and these equations can be presented in the form
\begin{equation}\label{f1f2}
\dfrac{\partial_1 f}f=\mathcal{F}_1(h_{ab},\omega^a{}_\gamma,\Phi_a),\qquad
\dfrac{\partial_2 f}f=\mathcal{F}_2(h_{ab},\omega^a{}_\gamma,\Phi_a),
\end{equation}
where the explicit expressions for $\mathcal{F}_1$ and $\mathcal{F}_2$ in terms  of the components of $\mathcal{F}_{\mu\nu}$ defined in (\ref{FXstructure}), with the notations $\alpha_1=\partial_1\alpha$ and $\alpha_2=\partial_2\alpha$, are
\[
\mathcal{F}_1=\dfrac{\alpha_1(\mathcal{F}_{11}+\epsilon \mathcal{F}_{22})-2\epsilon\alpha_2 \mathcal{F}_{12}}{2(\alpha_1^2-\epsilon\alpha_2^2)},\qquad
\mathcal{F}_2=\dfrac{-\alpha_2(\mathcal{F}_{11}+\epsilon \mathcal{F}_{22})+2\alpha_1 \mathcal{F}_{12}}{2(\alpha_1^2-\epsilon\alpha_2^2)}.
\]
Instead of the third of the mentioned above equations, it is more convenient to consider the equation for a Ricci scalar $R-4\Lambda=0$ which, multiplied by $(-f\alpha)$, reads as
\[
\alpha\partial_\gamma\Bigl(\dfrac{f^\gamma}f\Bigr)+2\, \eta^{\gamma\delta}\partial_\gamma\partial_\delta\alpha-\dfrac{f}{2\alpha^3}(L_c h^{cd} L_d)+\dfrac{\epsilon}{4\alpha} \eta^{\gamma\delta}\partial_\gamma h_c{}^d \partial_\delta h_d{}^c+4\alpha f\Lambda=0
\]
It is remarkable that after a very long calculations it can be shown that this equation as well as the compatibility condition $\partial_1 \mathcal{F}_2-\partial_2 \mathcal{F}_1=0$ of a pair of equations (\ref{f1f2}) are satisfied identically,  provided all field variables satisfy the constraint equations (\ref{Phimu}), (\ref{omegas}) and (\ref{f1f2}) as well as the self-dual Maxwell equations (\ref{sdMaxwell}) and the last subsystem in (\ref{RRR}), i.e. $R_a{}^b=8\pi T_a{}^b+\Lambda\delta_a{}^b$ which structure is considered just below.

\subsubsection{Self-dual form of symmetry-reduced Einstein - Maxwell equations}
\medskip

Similarly to the self-dual Maxwell equations (\ref{sdMaxwell}),  following Kinnersley \cite{Kinnersley:1973}, one can express the projections of the Ricci tensor $R^i{}_{(a)}$ in terms of self-dual bivectors ${\overset + K}{}_{ij(a)}$:
\begin{equation}\label{Rprojection}
R^i{}_{(a)}=\dfrac i 4\varepsilon^{ijkl}\nabla_j {\overset + K}{}_{kl (a)}\,,\qquad {\overset + K}{}_{ij(a)}\equiv K_{ij(a)}+i {\overset \ast K}{}_{ij(a)},\quad K_{ij(a)}\equiv \nabla_i \xi_{j(a)}-\nabla_j \xi_{i(a)}
\end{equation}
where the suffix in parenthesis numerates the Killing vector fields (\ref{Killings}), and the dual bivectors ${\overset \ast K}{}_{ij(a)}$ are defined similarly to a dual Maxwell tensor  (\ref{dualMaxwell}).
For the corresponding projections of the energy-momentum tensor we obtain
\begin{equation}\label{Tprojections}
T^i{}_{(a)}=\varepsilon^{ijkl}\nabla_j {\overset + S}{}_{kl (a)}+\dfrac 1{8\pi}{\overset + F}{}^{ij}(\mathcal{L}{}_{\xi_{(a)}}\overline{\Phi}_j)
\,,\qquad {\overset + S}{}_{ij(a)}=-\dfrac i{16\pi} \overline{\Phi}{}_{(a)} {\overset + F}{}_{ij},\qquad \Phi_{(a)}\equiv \Phi_k\xi^k{}_{(a)},
\end{equation}
where a bar means complex conjugation, bivectors ${\overset + S}{}_{kl (a)}$ with $a=3,4$ are self-dual and $\mathcal{L}{}_{\xi_{(a)}}$ means the Lie derivatives with respect to the Killing vector fields $\xi_{(a)}$. Therefore,  the second and third subsystems of Einstein - Maxwell equations (\ref{RRR}) take  the form
\begin{equation}\label{Hbivectors}
\varepsilon^{ijkl}\nabla_j {\overset + H}{}_{kl (a)}=-4 i {\overset + F}{}^{ij}(\mathcal{L}{}_{\xi_{(a)}}\overline{\Phi}_j)-4 i \Lambda\delta^i{}_{(a)},\quad\text{where}\quad {\overset + H}{}_{ik (a)}\equiv {\overset + K}{}_{ik (a)}+32 i\pi{\overset + S}{}_{ik(a)}.
\end{equation}
For each Killing vector field, $ {\overset + H}{}_{ij (a)}$ are self-dual bivectors:
\begin{equation}\label{sdH}
 {\overset + H}{}^{ij}{}_{(a)}=\dfrac i2 \varepsilon^{ijkl} {\overset + H}{}_{kl (a)}
\end{equation}

In contrast to the case of self-dual form of Maxwell equations (\ref{sdMaxwell}), the selfdual bivectors ${\overset + H}{}^{ij}{}_{(a)}$ do not possess the potentials because the right hand side of the equation (\ref{Hbivectors}) does not vanish, besides the special cases ($\Lambda=0$, $e_o=0$) or  ($\Lambda=0$, $\Phi_\mu=0$).
However, in general some matrix potentials associated with self-dual bivectors ${\overset + H}{}^{ij}{}_{(a)}$ can be constructed and this leads to some (more complicate) self-dual form of dynamical equations for gravitational field. To obtain these equations, we consider  the components of self-dual bivectors defined above, in the coordinates adapted for $\mathcal{G}$-symmetry of space-time geometry and fields. In these coordinates and our notations  described earlier in this paper, the expressions (\ref{Rprojection}) for Ricci tensor components read as
\begin{equation}\label{Rabcomponents}
R^\mu{}_a=\dfrac{i}{4 f\alpha}\varepsilon^{\mu\gamma}\epsilon^{cd} \partial_\gamma{\overset + K}{}_{c d a},\qquad
R^b{}_a=\dfrac{i}{2 f\alpha}\varepsilon^{\gamma\delta}\epsilon^{bc} \partial_\gamma{\overset + K}{}_{\delta c a},
\end{equation}
where we omit the parenthesis in the last suffix  which numerates the Killing vector fields. The components of the bivectors ${\overset + K}{}_{ij(a)}$ possess the expressions
\begin{equation}\label{Kcomponents}
\begin{array}{l}
{\overset + K}{}_{\mu a (b)}=\partial_\mu h_{ab}+i\alpha^{-1}\varepsilon_\mu{}^\nu h_a{}^c\partial_\nu h_{c b}
+i\epsilon \epsilon_{ca}\omega^c{}_\mu L_b\\[1ex]
{\overset + K}{}_{a b (c)}=i\epsilon \epsilon_{ab} L_c,\qquad
L_a\equiv \dfrac {\alpha}f h_{ac}\mathcal{T}^c
\end{array}
\end{equation}
The Einstein - Maxwell equations corresponding to the first part of Ricci tensor components (\ref{Rabcomponents}) were considered in the previous section where it was shown that these equations lead to constraint equations for  metric functions $\omega^a{}_\mu$ and the explicit expressions (\ref{omegas}) for the functions $L_a$. Therefore, we have to consider now the Einstein - Maxwell equations corresponding to the second part of equations (\ref{Rabcomponents}). Using the explicit expressions for components of the energy-momentum tensor of electromagnetic field (\ref{TmunuTmuaTab}), the constraint equations (\ref{Phimu}) and self-dual Maxwell equations (\ref{sdMaxwelleqs}), we can write these equations as
\begin{equation}\label{Keqs}
\varepsilon^{\gamma\delta}\partial_\gamma\bigl(
{\overset + K}{}_{\delta a}{}^b+2\overline{\Phi}{}^b \partial_\delta\Phi_a+2 \overline{e}{}_o\Phi_\delta\delta_a{}^b
\bigr)=2 i f\alpha\Lambda\delta_a{}^b
\end{equation}
where the index $b$ was rised as a spinor index: ${\overset + K}{}_{\delta a}{}^b=-{\overset + K}{}_{\delta a c}\epsilon^{cb}$. Going further, we split this matrix equation into its "trace" and "traceless" parts. The first of them leads to a dynamical equation for $\alpha$:
\begin{equation}\label{trace}
\begin{array}{l}
\eta^{\gamma\delta}\partial_\gamma\partial_\delta\alpha+\dfrac f{2 \alpha^3} (L_c h^{cd} L_d)+
2 f\bigl(\alpha\Lambda+ \dfrac{\epsilon e_o \overline{e}_o}{\alpha}\bigr)=0.
\end{array}
\end{equation}
while the traceless part of (\ref{Keqs}) can be presented in the form
\begin{equation}\label{Nequations}
\varepsilon^{\gamma\delta}\partial_\gamma\bigl[
N_{\delta a}{}^b+i\epsilon \epsilon_{ca}\omega^c{}_\mu \epsilon^{b d}(\ell_d-2 i \epsilon \overline{e}_o \Phi{}_d)+
Z_\delta \delta_a{}^b\bigr]=0,
\end{equation}
where we introduced the notations ($M_{\mu a}=\epsilon_{ca}M_\mu{}^c$ and $M_\mu{}^a$ were defined in (\ref{Mamu})):
\begin{equation}\label{ZNequations}
N_{\mu a}{}^b=\partial_\mu h_a{}^b+i\alpha^{-1}\varepsilon_\mu{}^\nu h_a{}^c\partial_\nu h_c{}^b+2 \overline{\Phi}^b M_{\mu a}
\qquad Z_\mu\equiv i\epsilon\varepsilon_\mu{}^\gamma\partial_\gamma \alpha+\dfrac i2 \epsilon\omega^c{}_\mu L_c-\overline{\Phi}{}^c\partial_\mu\Phi_c.
\end{equation}
Taking into account the self-dual form of Maxwell equations (\ref{sdMaxwelleqs}), it is easy to see that the components $N_{\mu a}{}^b$  and $M_{\mu a}$ satisfy the "two-dimensional" duality relations:
\begin{equation}\label{2d-duality}
\begin{array}{l}
N_{\mu a}{}^b=i\alpha^{-1}\varepsilon_\mu{}^\nu h_a{}^c N_{\nu c}{}^b\\[1ex]
M_{\mu a}=i\alpha^{-1}\varepsilon_\mu{}^\nu h_a{}^c M_{\nu c}.
\end{array}
\end{equation}
These equations would coincide with the duality equations which were constructed by Kinnersley \cite{Kinnersley:1977} for electrovacuum, provided $N_{\mu a}{}^b$  and $M_{\mu a}$ would have the potentials, but this is not so
in a more general case considered here. However, the equations (\ref{Nequations}) mean that the expression in square brackets possess some matrix potential $\widehat{H}_a{}^b$ so that
\begin{equation}\label{Hhat}
\partial_\mu \widehat{H}_a{}^b =N_{\mu a}{}^b+i\epsilon\omega_{a\mu}(\ell^b-2 i \epsilon \overline{e}_o \Phi^b)+
Z_\mu \delta_a{}^b.
\end{equation}
where $\omega_{a\mu}=\epsilon_{ca}\omega^c{}_\mu$ and $\ell^b=\epsilon^{bc}\ell_c$.
Therefore, we can express the components of $N_{\mu a}{}^b$ in terms of the components of this potential as
\begin{equation}\label{Hhat1}
N_{\mu a}{}^b=\partial_\mu \widehat{H}_a{}^b-i\epsilon \omega_{a \mu} (\ell^b-2 i \epsilon \overline{e}_o \Phi^b)-
Z_\mu \delta_a{}^b
\end{equation}
Substitution of these expressions into the duality relations (\ref{2d-duality}) leads to some modification of Kinnesrley self-dual equations for gravitational field in our more general case in which the non-dynamical degrees of freedom do not vanish.\footnote{As we shall see below, the matrix potential $\widehat{H}_a{}^b$ in the limit of vanishing of non-dynamcal degrees of freedom of gravitational and electromagnetic fields, does not coincide with the known Kinnersley's matrix potential $H_a{}^b$ but differes from this by some additional terms. Therefore, we say here not about a generalized potential, but about a modified one.} These modified Kinnersley's equations for the dynamical variables $h_{ab}$ and $\Phi_a$ takes the form
\begin{equation}\label{MKinnequations}
\begin{array}{l}
\partial_\mu \widehat{H}_a{}^b-Z_{\mu a}{}^b=
i\alpha^{-1}\varepsilon_\mu{}^\nu h_a{}^c (\partial_\nu \widehat{H}_c{}^b-Z_{\nu c}{}^b),\qquad Z_{\mu a}{}^b\equiv i\epsilon \omega_{a \mu} (\ell^b-2 i \epsilon \overline{e}_o \Phi^b)+
Z_\mu \delta_a{}^b,\\[1ex]
\partial_\mu\Phi_a-e_o\omega_{a\mu}=i\alpha^{-1}\varepsilon_\mu{}^\nu h_a{}^c (\partial_\nu\Phi_c-e_o\omega_{c\nu}),\\[1ex]
\partial_\mu \widehat{H}_a{}^b =\partial_\mu h_a{}^b+i\alpha^{-1}\varepsilon_\mu{}^\nu h_a{}^c\partial_\nu h_c{}^b+2 \overline{\Phi}^b (\partial_\mu\Phi_a-e_o\omega_{a\mu})+Z_{\mu a}{}^b.
\end{array}
\end{equation}
Unfortunately, these duality equations and the dynamical  equation (\ref{trace}) for the function $\alpha$ do not represent a closed system
of dynamical equations for the dynamical variables $h_{ab}$ and $\Phi_a$. The non-dynamical degrees of freedom and the corresponding constants -- real $\ell_a$ and complex $e_o$ also enter the equations (\ref{trace}) and (\ref{MKinnequations}). In this general case, the decoupling of the equations does not take place and the complete set of Einstein - Maxwell field equations includes, besides the dynamical equation (\ref{MKinnequations}) for $h_{ab}$ and $\Phi_a$ and the equation (\ref{trace}) for the function $\alpha$, also the constraint equations (\ref{omegas}) for  $\omega^a{}_\mu$ and (\ref{f1f2}) for the conformal factor $f$.

\subsubsection{Dynamical degrees of freedom and dynamical equations}
\medskip

In this subsection we consider the second order equations which arise from $\mathcal{G}_2$-symmetry reduced Einstein - Maxwell equations for the components of metric $h_{ab}$ and complex electro\-mag\-netic potential $\Phi_a$. Though in general case considered in this paper the dynamical equations for these field components do not decouple
from the equations for other field components, we call here these field variables as the dynamical degrees of freedom because in the most important subcases, in which the other degrees of freedom of fields vanish and the field equations become integrable, just these field components play the role of dynamical variables for  gravitational and electromagnetic fields.
The equations (\ref{MKinnequations}) imply the second order dynamical equations for $h_{ab}$ and $\Phi_a$ which can be presented in a symmetric form
\begin{equation}\label{hdynamics}
\left\{
\begin{array}{l}
\eta^{\mu\nu}\partial_\mu\bigl[\alpha^{-1}(\partial_\nu h_{ac}) h^{cb}\bigr]+i\epsilon\varepsilon^{\mu\nu}\bigl( M_{\mu a}\overline{M}_\nu{}^b- \overline{M}{}_{\mu a} M_\nu{}^b \bigr)\\[1ex]
\qquad\qquad\qquad\qquad+\dfrac{\epsilon f}{\alpha^3} L_a L_c h^{cb} +2 f\bigl(\epsilon\alpha\Lambda+ \dfrac{ e_o \overline{e}_o}{\alpha}\bigr)\delta_a{}^b=0,\\[1ex]
\eta^{\mu\nu}\partial_\mu\bigl[\alpha^{-1}M_{\nu c} h^{cb}\bigr]-i e_o\alpha^{-3} f h^{bc} L_c=0.
\end{array}\right.
\end{equation}
where $L_a$ and $M_\gamma{}^a$ were defined in (\ref{omegas}) and (\ref{Mamu}) respectively.
The trace of the left hand side of the first equation in (\ref{hdynamics}), leads again to  dynamical equation (\ref{trace}) for $\alpha$:
\[
\begin{array}{l}
\eta^{\gamma\delta}\partial_\gamma\partial_\delta\alpha+\dfrac f{2 \alpha^3} (L_c h^{cd} L_d)+
2 f\bigl(\alpha\Lambda+ \dfrac{\epsilon e_o \overline{e}_o}{\alpha}\bigr)=0.
\end{array}
\]
Using this equation and introducing the new matrix variable $k_a{}^b$ such that
\begin{equation}\label{kproperties}
k_a{}^b\equiv \alpha^{-1} h_a{}^b,\qquad k_c{}^c=0,\qquad k_a{}^c k_c{}^b=-\epsilon\delta_a{}^b,\qquad\det\Vert k_a{}^b\Vert=-\epsilon,
\end{equation}
the equations (\ref{hdynamics}) can be simplified and presented in the "tracelrss" form
\begin{equation}\label{kLMdynamics}
\left\{
\begin{array}{l}
\eta^{\gamma\delta}\partial_\gamma\bigl[\alpha (\partial_\delta  k_a{}^c )k_c{}^b\bigr]+i\epsilon\varepsilon^{\gamma\delta}\bigl(\!\overline{M}{}_{\gamma a} M_\delta{}^b-\! M_{\gamma a} \overline{M}_\delta{}^b\bigr)\\[1ex]
\qquad\qquad\qquad\qquad
-\dfrac{\epsilon f}{\alpha^2}\bigl[L_a L_c k^{cb} -\dfrac 12 (L_c k^{cd} L_d)\delta^b{}_a\bigr]=0,\\[2ex]
\eta^{\mu\nu}\partial_\mu\bigl[M_{\nu c} k^{cb}\bigr]-i e_o\alpha^{-2} f k^{bc} L_c=0.
\end{array}\right.
\end{equation}
It is interesting to note here that the equations (\ref{hdynamics}) or the equations (\ref{kLMdynamics}) do not represent a closed systems for dynamical variables $h_{ab}$ and $\Phi_a$ or $k_{ab}$ and $\Phi_a$ respectively because the non-dynamical variables $f$ and $\omega^a{}_\mu$ (through the structure of $M_\mu{}^a$) also enter these equations. Having the aim to decouple the dynamical equations for $h_{ab}$ and $\Phi_a$ from the other (constraint) equations for other (non-dynamical) degrees of freedom, we could solve easily the equation (\ref{trace}) with respect to $f$ and thus to express it in terms of $h_{ab}$ and $\Phi_a$ only.
However, substitution of this expression for $f$ into the dynamical equations does not solve completely the problem of decoupling of these equations. Besides that, we have to understand, if some constraints on the field variables can arise if we substitute this expression for $f$ into the constraint equations (\ref{f1f2}). Nontheless, despite of the absence in general of decoupling of the equations (\ref{hdynamics}) as the equations for $h_{ab}$ and $\Phi_a$ from the constraint equations for $\omega^a{}_\mu$ and $f$, we call here the field variables $h_{ab}$ and $\Phi_a$ as dynamical variables and the equations (\ref{hdynamics}) as dynamical equations because in the most important subcase in which the constants $\ell_a$ and $e_o$ vanish, such decoupling take place and leads to the integrability of the dynamical equations (\ref{hdynamics}).  To understand what happens in the other  cases in which such decoupling do not take place, we restrict our considerations by some most interesting particular cases which illustrate this complicated structure of the equations.

\subsection{Classes of fields with non-dynamical degrees of freedom}
As it was shown in the previous section, in general class of electrovacuum spacetimes which metrics admit the Abelian isometry group $\mathcal{G}_2$, gravitational and electromagnetic fields can possess (besides the well known dynamical degrees of freedom) some non-dynamical degrees of freedom. The gravitational and electromagnetic field components corresponding to these non-dynamical degrees of freedom can be determined completely as the solutions of the constraint equations which include a set of arbitrary constants which characterise these degrees of freedom. This set of constants includes the components of constant real two-dimensional vector $\ell_a$ which non-zero components lead to such property of space-time geometry that the orbits of the isometry group $\mathcal{G}_2$ are not $2$-surface orthogonal and therefore,
the space-time metric can not possess a block-diagonal structure, and the complex constant $e_o$ which possess the electromagnetic nature. Its non-zero value leads to more complicate structure of electromagnetic field. A striking similarity of the structure of the field equations for these electrovacuum fields with non-dynamical degrees of freedom and of elecrovacuum field equations with $\mathcal{G}_2$ isometry group and non-vanishing cosmological constant made it reasonable to include in our considerations the corresponding class of fields with non-vanishing cosmological constant as one more non-dynamical degree of freedom of gravitational field.

\subsubsection{Vacuum fields with cosmological constant ($\Lambda\ne 0$)}
\medskip

In this subsection we consider physically important subclass of the described above class of spacetimes.  This is a class of pure vacuum metrics with cosmological constant and with the Abelian isometry group $\mathcal{G}_2$. For this subclass, we choose for simplicity
\begin{equation}\label{Lambdavacuum}
\ell_a=0,\qquad e_o=0,\qquad \Phi_a=0,\qquad \Lambda\ne 0.
\end{equation}
With these conditions, the equation (\ref{hdynamics}) can be presented in  the form
\begin{equation}\label{hdynamics1}
\begin{array}{l}
\eta^{\mu\nu}\partial_\mu\bigl[\alpha^{-1} h_a{}^c\partial_\nu h_c{}^b\bigr]+2\epsilon\alpha f \Lambda\delta_a{}^b=0
\end{array}
\end{equation}
and, taking the trace of (\ref{hdynamics1}) or using the condition (\ref{Lambdavacuum}) in (\ref{trace}), we obtain
\begin{equation}\label{fexpression}
f=-\dfrac{\eta^{\gamma\delta}\partial_\gamma\partial_\delta\alpha}{2
\alpha\Lambda}
\end{equation}
Substituting this expression into (\ref{hdynamics1}) we obtain a closed system of dynamical equations for a matrix $k_a{}^b\equiv \alpha^{-1} h_a{}^b$ which algebraic properties was already mentioned in (\ref{kproperties}):
\begin{equation}\label{kdynamics}
\eta^{\mu\nu}\partial_\mu\bigl[\alpha\, k_a{}^c\partial_\nu k_c{}^b\bigr]=0,\qquad k_c{}^c=0,\qquad \det\Vert k_a{}^b\Vert=\epsilon,\qquad
k_a{}^c k_c{}^b=-\epsilon \delta_a{}^b.
\end{equation}

\paragraph{\rm\textit{\underline{\textit{Dynamical equations in terms of scalar functions.}}}}
Using (\ref{fexpression}) and parametrization
\begin{equation}\label{Parametrization}
\Vert g_{ab}\Vert\equiv \Vert h_{ab}\Vert =
\epsilon_0\begin{pmatrix}
H&H\Omega\\
H \Omega&H\Omega^2+\epsilon\alpha^2 H^{-1}
\end{pmatrix},
\end{equation}
where $\epsilon_0=\pm 1$ is the sign of $g_{33}$ and the functions $\alpha>0$, $H > 0$ and $\Omega=g_{34}/g_{33}$, the  equations (\ref{hdynamics1}) can be reduced to a pair of  equations for two scalar functions $H$ and $\Omega$:
\begin{equation}\label{HOmegaequations}
\left\{\begin{array}{l}
\eta^{\mu\nu}\bigl(
H_{\mu\nu}+\dfrac{\alpha_\mu}\alpha H_\nu-\dfrac{\alpha_{\mu\nu}}\alpha H-\dfrac{H_\mu H_\nu}{H} -\epsilon\dfrac{H^3}{\alpha^2}\Omega_\mu\Omega_\nu\bigr)=0,\\[2ex]
\eta^{\mu\nu}\bigl(\Omega_{\mu\nu}-\dfrac{\alpha_\mu}\alpha \Omega_\nu+2\dfrac{H_\mu}H \Omega_\nu\bigr)=0.
\end{array}\right.
\end{equation}
Here and below, the greek letters as well as the numbers $1$ and $2$ in the suffices of scalar functions mean the partial derivatives with respect to the coordinates $x^\mu=\{x^1,x^2\}$.

\paragraph{\rm\textit{\underline{\textit{Generalization of the Ernst equations.}}}}
Sometimes it is convenient, similarly to pure vacuum case \cite{Ernst:1968}, to rewrite the second of the equations (\ref{HOmegaequations}) in a different form which allows to introduce the potential $\phi$:
\[\eta^{\mu\nu}\partial_\mu(\alpha^{-1}H^2\partial_\nu\Omega)=0\quad \Rightarrow \quad \partial_\mu \Omega=\alpha H^{-2}\varepsilon_\mu{}^\nu\partial_\nu\phi.
\]
Using this potential, the equations (\ref{HOmegaequations}) can be presented in the form
\begin{equation}\label{Hphiequations}
\left\{\begin{array}{l}
\eta^{\mu\nu}\bigl(
H_{\mu\nu}+\dfrac{\alpha_\mu}\alpha H_\nu-\dfrac{\alpha_{\mu\nu}}\alpha H-\dfrac{H_\mu H_\nu}{H} +\dfrac 1 H\phi_\mu\phi_\nu\bigr)=0,\\[2ex]
\eta^{\mu\nu}\bigl(\phi_{\mu\nu}+\dfrac{\alpha_\mu}\alpha \phi_\nu-2\dfrac{H_\mu}H \phi_\nu\bigr)=0.
\end{array}\right.
\end{equation}
Following again pure vacuum procedure \cite{Ernst:1968} we can combine these equations into the Ernst-like equation for a complex potential $\mathcal{E}\equiv H+i\phi$ which takes the form
\begin{equation}\label{Ernsteqs}
(\text{Re\,} \mathcal{E})\,\eta^{\mu\nu}\bigl(
\mathcal{E}_{\mu\nu}+\dfrac{\alpha_\mu}\alpha \mathcal{E}_\nu-\dfrac{\alpha_{\mu\nu}}\alpha \text{Re\,} \mathcal{E}\bigr) -\eta^{\mu\nu}\mathcal{E}_\mu \mathcal{E}_\nu =0
\end{equation}
where the only difference with vacuum Ernst equation is the presence of the last term in the parenthesis, while for vacuum vith $\Lambda=0$ this term vanishes because in this case $\eta^{\mu\nu}\alpha_{\mu\nu}=0$.

\paragraph{\rm\textit{\underline{\textit{Complete system of field equations.}}}}
Another compact form of the equations (\ref{Hphiequations}) arises if we use a new unknown function $\psi$ such that $H=\alpha e^\psi$. Then we obtain
\begin{equation}\label{psiphieqs}
\left\{\begin{array}{l}
\eta^{\mu\nu}(\psi_{\mu\nu}+\dfrac{\alpha_\mu}\alpha \psi_\nu+\alpha^{-2} e^{-2\psi}\phi_\mu\phi_\nu)=0, \\[1ex]
\eta^{\mu\nu}(\phi_{\mu\nu}-\dfrac{\alpha_\mu}\alpha \phi_\nu-2\psi_\mu\phi_\nu)=0.
\end{array}\right.
\end{equation}
Besides the equations  (\ref{psiphieqs}) or, equivalently, the equation (\ref{Ernsteqs}), some supplement equations for the functions $\psi$, $\phi$ and $\alpha$ arise from the equations (\ref{f1f2}) and (\ref{fexpression}). These supplement equations possess the structures
\begin{equation}\label{supplementeqs}
\left\{\begin{array}{l}
\psi_1^2+\epsilon\psi_2^2+\alpha^{-2} e^{-2\psi}(\phi_1^2+\epsilon\phi_2^2)=\mathcal{K}_1 \\[1ex]
2\psi_1\psi_2+2\alpha^{-2} e^{-2\psi}\phi_1\phi_2=\mathcal{K}_2
\end{array}\right.
\end{equation}
where the right hand sides $\mathcal{K}_1$, $\mathcal{K}_2$ are determined completely by the function $\alpha(x^1,x^2)$:
\begin{equation}\label{K1K2I}
\left.\begin{array}{l}
\mathcal{K}^{(I)}_1=\dfrac{2 B}{\sqrt{\alpha} A}\left[\alpha_1\bigl(\dfrac{A}{\sqrt{\alpha}B}\bigr)_1+\epsilon \alpha_2\bigl(\dfrac{A}{\sqrt{\alpha}B}\bigr)_2\right]+2(\alpha_1^2 +\epsilon \alpha_2^2)\dfrac{A}{\alpha B},\\[3ex]
\mathcal{K}^{(I)}_2=\dfrac{2 B}{\sqrt{\alpha} A}\left[\alpha_1\bigl(\dfrac{A}{\sqrt{\alpha} B}\bigr)_2+ \alpha_2\bigl(\dfrac{A}{\sqrt{\alpha} B}\bigr)_1\right]+4\alpha_1 \alpha_2 \dfrac{A}{\alpha B},
\end{array}\quad\right\Vert\quad
\begin{array}{l}
A=\alpha_{11}-\epsilon\alpha_{22},\\[1ex]
B=\alpha_1^2-\epsilon\alpha_2^2.
\end{array}
\end{equation}
It is worth noting here that a complete system of field  equations which we obtain here, i.e. the equations (\ref{psiphieqs}) and (\ref{supplementeqs}) together with (\ref{K1K2I}) does not admit a direct passage to the limit $\Lambda=0$ and to the corresponding condition $A=0$, because we use the expression (\ref{fexpression}) for the conformal factor where it was assumed that $\Lambda\ne 0$.

 A complete system of field  equations (\ref{psiphieqs}) and (\ref{supplementeqs}) together with (\ref{K1K2I}) possess rather complicate structure which does not allow, for example, to analyse the general structure of its space of solutions. However, we can clarify the situation a bit more, if we restrict our considerations by a class of fields with diagonal metrics -- static fields ($\epsilon=-1$) or, e.g., plane waves with linear polarization ($\epsilon=1$).

\paragraph{\rm\textit{\underline{\textit{The fields with diagonal metrics.}}}} For the class of fields with diagonal metrics we have
\[\Omega=0,\qquad \phi=0\]
and the equations (\ref{psiphieqs}), (\ref{supplementeqs}) and (\ref{K1K2I}) reduces by the substitution $H=\alpha e^\psi$ to the equations
\begin{equation}\label{psieqsI}
\left\{\begin{array}{l}
\psi_{11}-\epsilon\psi_{22}+\dfrac{\alpha_1}\alpha \psi_1- \epsilon \dfrac{\alpha_2}\alpha \psi_2=0,\\[1ex]
\psi_1^2+\epsilon\psi_2^2=\mathcal{K}_1 \\[1ex]
2\psi_1\psi_2=\mathcal{K}_2
\end{array}\right.
\end{equation}
where the functions $\mathcal{K}_1$ and $\mathcal{K}_2$ are determined in terms of the function $\alpha$ by the expressions
(\ref{K1K2I}), i.e. $\mathcal{K}_1=\mathcal{K}^{(I)}_1$ and $\mathcal{K}_2=\mathcal{K}^{(I)}_2$.
The last two equations in (\ref{psieqsI}) can be solved directly with respect to $\psi_1$ and $\psi_2$. This leads to the following solutions:
\begin{equation}\label{psiK1K2I}
\psi_1=\pm\dfrac 1{\sqrt{2}} \sqrt{\mathcal{K}_1+\sqrt{\mathcal{K}_1^2-\epsilon \mathcal{K}_2^2}},\qquad
\psi_2=\pm\dfrac 1{\sqrt{2}}\,\dfrac{\mathcal{K}_2} {\sqrt{\mathcal{K}_1+\sqrt{\mathcal{K}_1^2-\epsilon \mathcal{K}_2^2}}}.
\end{equation}
where one should choose only the upper signs or only the lower signs.
(Another solution arise if we interchange the expressions for $\psi_1$ and $\psi_2$.) Besides that, as it is easy to see, an aditional condition arises from (\ref{psieqsI}). This condition is
\[\mathcal{K}_1 >0\quad\text{for}\quad \epsilon=1.
\]
The expressions (\ref{psiK1K2I}) for $\psi_1$ and $\psi_2$  give rise also to a pair of equations which should be satisfied by the functions $\mathcal{K}_1$ and $\mathcal{K}_2$. These equations  arise from the first equation in (\ref{psieqsI}) and the compatibility condition
\[\partial_1(\psi_1)-\epsilon \partial_2(\psi_2)+\dfrac{\alpha_1}\alpha \psi_1- \epsilon \dfrac{\alpha_2}\alpha \psi_2=0,\qquad \partial_1\psi_2-\partial_2\psi_1=0.
\]
Using  (\ref{psiK1K2I}), these equations can be transformed into the equations for $\mathcal{K}_1$ and $\mathcal{K}_2$:
\begin{equation}\label{K1K2eqsI}
\left\{\begin{array}{l}
\partial_1 \mathcal{K}_2-\partial_2 \mathcal{K}_1 +\dfrac{\alpha_1}{\alpha} \mathcal{K}_2+\dfrac{\alpha_2}{\alpha}(-\mathcal{K}_1+\sqrt{\mathcal{K}_1^2-\epsilon \mathcal{K}_2^2})=0,\\[2ex]
\partial_2 \mathcal{K}_2-\epsilon\partial_1 \mathcal{K}_1 +\dfrac{\alpha_2}{\alpha} \mathcal{K}_2-\dfrac{\epsilon \alpha_1}{\alpha}(\mathcal{K}_1+\sqrt{\mathcal{K}_1^2-\epsilon \mathcal{K}_2^2})=0.
\end{array}\right.
\end{equation}
where $\mathcal{K}_1$ and $\mathcal{K}_2$ were expressed in (\ref{K1K2I}) in terms of the function $\alpha$ and its derivetives. However, it is easy to show that in (\ref{K1K2eqsI}) we have only one equation for $\alpha$, because certain linear combination of the equations (\ref{K1K2eqsI}), which does not include the square root, reduces to identity in view of (\ref{K1K2I}). Therefore, we can consider only one of the equations (\ref{K1K2eqsI}), e.g. the first one, which can be presented in the form
\begin{equation}\label{AlphaequationI}
\partial_2(\alpha \mathcal{K}_1)-\partial_1(\alpha \mathcal{K}_2)=\alpha_2 \sqrt{\mathcal{K}_1^2-\epsilon \mathcal{K}_2^2}Ў
\end{equation}
where $\mathcal{K}_1=\mathcal{K}^{(I)}_1$ and $\mathcal{K}_2=\mathcal{K}^{(I)}_2$ which were defined in (\ref{K1K2I}).
Taking "square" of this equation leads to a polynomial equation for $\alpha$ and its derivatives up to the fourth order. This polynom can be reduced a bit because it can be factorized and the multiplier $\alpha_2^2(\alpha_1^2-\epsilon\alpha_2^2)^4$ can be canceled. As a result, we obtain still rather complicate, highly nonlinear differential equation for $\alpha$ which left hand side is a polynom with respect to $\alpha$ and its derivatives of all orders up to the fourth one. This polynomial is homogeneous with respect to $\alpha$ and its degree of homogeneity is equal to $7$. We do not present here the explicit form of this polynomial differential equation  because it is rather long. Though a complicate structure of this equation does not allow to expect that this equation for $\alpha$ may occur to be completely integrable, its existence shows the structure of the space of solutions of the class of diagonal vacuum metrics with cosmological constant which possess the Abelian isometry group $\mathcal{G}_2$. Namely, the derived nonlinear fourth-order differential equation for $\alpha$ actually represent the only dynamical equation, which solutions determine completely all corresponding metric components using the relations derived above. However, it is necessary to mention, that there are some restrictions on the choice of the solutions of dynamical equation for $\alpha$. Besides an obvious condition $\alpha>0$, in the hyperbolic case ($\epsilon=1$) the condition  $\mathcal{K}_1>0$ should be satisfied, where $\mathcal{K}_1$ is determined in terms of $\alpha$ in (\ref{K1K2I}). Another condition for the choice of the solution of the mentioned above squared (polynomial) differential equation is that this solution should satisfy also the original equation (\ref{AlphaequationI}) which space of solutions includes  only "half" of the solution space of the squared equation.

\subsubsection{Vacuum fields with non-ortogonally-transitive isometry group  $\mathcal{G}_2$ ($\omega^a{}_\mu\ne 0$)}
\medskip

To consider the simplest subcase of general class of metrics with non-dynamical degrees of freedom discussed in this paper, in which the orbits of the isometry group $\mathcal{G}_2$ are not 2-surface-orthogonal and therefore, the metrics do not possess the $2\times 2$-block-diagonal form ($\omega^a{}_\mu\ne 0$), we chose the case of vacuum fields with $\mathcal{G}_2$-symmetry for which we put
\[\Lambda=0,\quad \Phi_a=0,\quad e_o=0,\quad \ell_a\ne0.
\]
Using a freedom of $SL(2,R)$ linear transformations of the coordinates $x^a=(x^3,x^4)$, we can reduce, without any loss of generality, the components of the constant vector $\ell_a$ to the simplest form and assume for simplicity that in the transformed coordinates the matrix $h_{ab}$ of metric components is diagonal. Thus, we put
\[\ell_a=\{\ell_o,\,0\},\qquad \Omega=0.\]
In this case, similarly to the previous case of vacuum fields with cosmological constant, we can find again the conformal factor $f$ from the equation (\ref{trace}):
\[f=-\dfrac{2\epsilon\epsilon_0\epsilon_1}{\ell_o^2} \alpha H A
\]
where the sign $\epsilon_0$ and the function $H$ arise from the parametrization (\ref{Parametrization}) for  $h_{ab}$. The functions $A$ and $B$ were defined in (\ref{K1K2I}).
It is very surprising that in the present case the general equations (\ref{hdynamics}) after a different sabstitution for $H=\alpha^2 e^\psi$ reduce to the same equations (\ref{psieqsI}),
but the expressions for  $\mathcal{K}_1$ and $\mathcal{K}_2$
differ from the expressions  (\ref{K1K2I}):
\begin{equation}\label{K1K2II}
\mathcal{K}^{(II)}_1=\mathcal{K}^{(I)}_1+7\,\dfrac{\alpha_1^2+ \epsilon\alpha_2^2}{\alpha^2},\qquad
\mathcal{K}^{(II)}_2=\mathcal{K}^{(I)}_2+14\,\dfrac{\alpha_1 \alpha_2}{\alpha^2}.
\end{equation}
Further, the basic equations for this class can be constructed using the same expressions (\ref{psieqsI}) -- (\ref{AlphaequationI}), where we have to use $\mathcal{K}_1=\mathcal{K}^{(II)}_1$ and $\mathcal{K}_2=\mathcal{K}^{(II)}_2$.
As a result, the polynomial equation for $\alpha$, which we obtain "squaring"  (\ref{AlphaequationI}), will differ from (\ref{AlphaequationI}) only by numerical coefficients.

\subsubsection{Electrovacuum fields with non-dynamical degrees of freedom ($e_o\ne 0$)}
\medskip
For simplicity, we consider here the class of elecromagnetic fields with non-dynamical electro\-mag\-netic degrees of freedom for which the metric $h_{ab}$ on the orbits of the isometry group is diagonal and besides that, the projections of complex electromagnetic potential on the orbits, the metric components which make the orbits to be not $2$-surface orthogonal and the cosmological constant vanish, i.e. for this class of fields
\[\Omega=0,\quad \Phi_a=0,\quad \ell_a=0,\quad \Lambda=0,\quad e_o\ne 0.
\]
In this case, similarly to the previous cases, we find the conformal factor $f$ from (\ref{trace}):
\[f=-\dfrac{\epsilon\epsilon_1 \alpha}{2 e_o \overline{e}_o} A
\]
It is remarkable that in this case the same substitution as in the case of presence of cosmological constant, i.e. $H=\alpha e^\psi$ leads to the same dynamical equations (\ref{psieqsI}) for $\psi$ and $\alpha$, in which, however, the expressions for $\mathcal{K}_1$ Ё $\mathcal{K}_2$, denoted here as  $\mathcal{K}^{(III)}_1$ and $\mathcal{K}^{(III)}_2$, differ from  (\ref{K1K2eqsI}) and are determined by the expressions
\begin{equation}\label{K1K2III}
\mathcal{K}^{(III)}_1=\mathcal{K}^{(I)}_1+4\,\dfrac{\alpha_1^2+ \epsilon\alpha_2^2}{\alpha^2},\qquad
\mathcal{K}^{(III)}_2=\mathcal{K}^{(I)}_2+8\,\dfrac{\alpha_1 \alpha_2}{\alpha^2}.
\end{equation}
Further reduction of these dynamical equations follows the same equations  (\ref{psieqsI}) -- (\ref{AlphaequationI}), where   $\mathcal{K}_1=\mathcal{K}^{(III)}_1$ and $\mathcal{K}_2=\mathcal{K}^{(III)}_2$. This leads again to the only dynamical equation for $\alpha$ which differs from the similar equations derived in two previous cases only by numerical coefficients.

\subsection{Examples of solutions with (I) $\Lambda\ne 0$, (II) $\omega^a{}_\mu\ne 0$, (III) $e_o\ne 0$}
\medskip
In this section, we describe a construction of (presumably) new families of  solutions of all three types (I), (II) and (III) and give explicitly the simplest examples.

The particular solutions for all three classes of fields considered above can be constructed in a unified way, if we begin with the system of equations for the functions $\psi$ and $\alpha$
\begin{equation}\label{psieqs}
\left\{\begin{array}{l}
\psi_{11}-\epsilon\psi_{22}+\dfrac{\alpha_1}\alpha \psi_1- \epsilon \dfrac{\alpha_2}\alpha \psi_2=0,\\[1ex]
\psi_1^2+\epsilon\psi_2^2=\mathcal{K}_1, \\[1ex]
2\psi_1\psi_2=\mathcal{K}_2,
\end{array}\right.
\end{equation}
which arises in the same form for all three classes of fields, but with different values of $\mathcal{K}_1$ and $\mathcal{K}_2$, defined respectively in (\ref{K1K2I}),  (\ref{K1K2II}) or (\ref{K1K2III}). However, introducing the parameter $n$, which takes the values $n=1$ for the class (I), $n=-6$ for the class (II) and $n=-3$ for the class (III), all expressions for $\mathcal{K}_1$ and $\mathcal{K}_2$ can be presented in a convenient general form
\begin{equation}\label{K1K2}
\begin{array}{l}
\mathcal{K}_1=\dfrac{2 B}{\sqrt{\alpha} A}\left[\alpha_1\bigl(\dfrac{A}{\sqrt{\alpha}B}\bigr)_1+\epsilon \alpha_2\bigl(\dfrac{A}{\sqrt{\alpha}B}\bigr)_2\right]+\dfrac{(\alpha_1^2 +\epsilon \alpha_2^2)}{\alpha}\Bigl(\dfrac{2 A}{B}+\dfrac{1-n}{\alpha}\Bigr),\\[3ex]
\mathcal{K}_2=\dfrac{2 B}{\sqrt{\alpha} A}\left[\alpha_1\bigl(\dfrac{A}{\sqrt{\alpha} B}\bigr)_2+ \alpha_2\bigl(\dfrac{A}{\sqrt{\alpha} B}\bigr)_1\right]+\dfrac{2\alpha_1 \alpha_2}{\alpha} \Bigl(\dfrac{2 A}{B}++\dfrac{1-n}{\alpha}\Bigr),
\end{array}
\end{equation}
where, as before, we use the notations $A=\alpha_{11}-\epsilon\alpha_{22}$ Ё $B=\alpha_1^2-\epsilon\alpha_2^2$.

\subsubsection{Construction of solutions for the classes of fields (I), (II), (III)}
To solve the equations derived above, we use a simple ansatz in which the unknown functions $\psi$ and $\alpha$ are assumed to be depending on a one unknown function only:
\begin{equation}\label{psialpha}
\psi=\psi(w),\qquad \alpha=\alpha(w),\qquad w=w(x^1,x^2).
\end{equation}
Substitution of this ansatz into the first of the equations (\ref{psieqs}) solves this equation provided the function $w$ satisfies the following linear equation and the functions $\psi(w)$ and $\alpha(w)$ satisfy the easily solvable relation
\begin{equation}\label{wequation}
w_{11}-\epsilon w_{22}=0,\qquad \psi^{\prime\prime}+\dfrac{\alpha^\prime}{\alpha}\psi^\prime=0\qquad \Rightarrow\qquad
\psi^\prime(w)=\dfrac{k_o}{\alpha(w)}.
\end{equation}
where $k_o$ is an arbitrary real constant. Then, we also obtain from  (\ref{psieqs})
\begin{equation}\label{K1K2ko}
\mathcal{K}_1=\dfrac{k_o^2}{\alpha^2(w)}(w_1^2+\epsilon w_2^2),\qquad \mathcal{K}_2=\dfrac{2 k_o^2}{\alpha^2(w)} w_1 w_2.
\end{equation}
Taking into account the linear equation (\ref{wequation}) for $w$, one finds that $\alpha(w)$ should satisfy an ordinary differential equation
\begin{equation}\label{alphaeqnkon}
2\alpha\bigl[\alpha^\prime\alpha^{\prime\prime\prime}- (\alpha^{\prime\prime})^2\bigr]-\bigl[k_o^2+n(\alpha^\prime)^2\bigr] \alpha^{\prime\prime}=0
\end{equation}
where the prime denotes a differentiation with respect to $w$. A standard sabstitution (the dot denotes a differentiation with respect to $\alpha$)
\begin{equation}\label{Pdef}
\alpha^\prime=P(\alpha),\qquad \alpha^{\prime\prime}=\dot{P} P,\qquad
\alpha^{\prime\prime\prime}=\ddot{P} P^2+\dot{P}^2 P,
\end{equation}
after a change of the independent variable $\alpha=e^x$ transforms the equation  (\ref{alphaeqnkon}) into
\begin{equation}\label{Peqn}
P_{xx}-\dfrac12 \bigl(\dfrac{k_o^2}{P^2}+n+2\bigr) P_x=0.
\end{equation}
If we introduce the notations for auxiliary constants (where $g_o$ is an  arbitrary real constant)
\begin{equation}\label{parameters}
h_{\pm}=\dfrac{P_{\pm}}{(n+2) R_o},\qquad P_{\pm}=g_o\pm R_o,\qquad  R_o=\sqrt{g_o^2+\dfrac{k_o^2}{n+2}},\qquad h_o=\dfrac{k_o}{(n+2)R_o}
\end{equation}
and set in the expressions for $P_\pm$ and $h_\pm$ respectively
\[(I):\quad n=1,\qquad (II):\quad n=-6,\qquad (III):\quad n=-3,
\]
the solution for the above equations can be presented in a parametric form
\begin{equation}\label{solutionI-II-III}
\alpha=\alpha_o\dfrac{(P-P_+)^{h_+}}{(P-P_-)^{h_-}},\qquad
e^\psi=e^{\psi_o}\left(\dfrac{P-P_+}{P-P_-}\right)^{h_o},\quad
w=\dfrac {2 \alpha_o}{n+2}\int\dfrac{(P-P_+)^{-1+h_+}}{(P-P_-)^{1+h_-}} dP,
\end{equation}
where $\alpha_o$ and $\psi_o$ are arbitrary real constants and the integral for $w$ can be expressed in terms of the hypergeometric function. For the conformal factor $f$ we have different expressions for different cases:
\begin{equation}\label{fI-II-III}
\begin{array}{l}
f^{(I)}=-\dfrac{3\epsilon_1}{4\alpha_o^2\Lambda}( w_1^2-\epsilon w_2^2)(P-P_+)^{1-2 h_+}
(P-P_-)^{1+2 h_-},\\[3ex]
f^{(II)}=\dfrac{4\epsilon\epsilon_0\epsilon_1\alpha_o^2 e^{\psi_o}}{\ell_o^2} (w_1^2-\epsilon w_2^2)(P-P_+)^{1+h_o+2h_+}
(P-P_-)^{1-h_o-2h_-},\\[3ex]
f^{(III)}=\dfrac{\epsilon\epsilon_1}{4 e_o\overline{e}_o} ( w_1^2-\epsilon w_2^2)(P-P_+)(P-P_-).
\end{array}
\end{equation}
where we have to choose in the expressions for $P_\pm$ and $h_\pm$ the corresponding values $n=1$, $n=-6$ and $n=-3$. The choice of the signs  $\epsilon$, $\epsilon_0$, $\epsilon_1$  and of the solution $w$ of the linear equation (\ref{wequation}) should satisfy the only condition $f>0$. Besides that, in the case (II) the equations for calculation of metric functions $\omega^a{}_\mu$ (which presence is an obstacle for the existence of 2-surfaces orthogonal to the isometry group orbits) take the form
\begin{equation}\label{omegasolution}
\partial_1\omega^a{}_2- \partial_2\omega^a{}_1=\dfrac{4\epsilon\epsilon_1} {\alpha_o \ell_o} ( w_1^2-\epsilon w_2^2)(P-P_+)^{1-h_+}
(P-P_-)^{1+h_-}\delta^a_3,
\end{equation}
and in the case (III) we obtain the similar equation for calculation of the components $\Phi_\mu$ of complex electromagnetic potential
\begin{equation}\label{Phisolution}
\partial_1\Phi_2- \partial_2\Phi_1=-\dfrac{i \epsilon_1}{4\alpha_o \overline{e}_o} ( w_1^2-\epsilon w_2^2)(P-P_+)^{1-h_+}
(P-P_-)^{1+h_-},
\end{equation}
where in the expressions for $P_\pm$ and $h_\pm$ in (\ref{omegasolution}) we should choose $n=-6$, and in (\ref{Phisolution}) we should choose $n=-3$.
Each of these two equations can be solved if we use a gauge transformations $\omega^a{}_\mu \to \omega^a{}_\mu +\partial_\mu k^a$ and $\Phi_\mu \to \Phi_\mu+\partial_\mu \phi$ ив жаиба $k^a$ Ё $\phi$ are arbitrary functions. This allows to simplify the components $\omega^a{}_\mu$ and $\Phi_\mu$, choosing one of the components $\omega^3{}_\mu$ and one of the components $\Phi_\mu$, as well as well as both components $\omega^4{}_\mu$ equal to zero. Then, nonvanishing components among $\omega^3{}_\mu$ and $\Phi_\mu$ can be expressed in quadratures (see the examples given below).

It is important to note here that though the constructed above solutions depend on arbitrary choosing solution $w(x^1,x^2)$ of the linear equation (\ref{wequation}) and the constants of integration $k_o$ and $g_o$, these solutions represent only finite-parametric families, while the arbitrary function can be excluded from the solutions by an appropriate coordinate transformation.

\paragraph{\rm\textit{\underline{\textit{In the hyperbolic case ($\epsilon=1$),}}}} we may choose $\epsilon_1=-1$ and $(x^1,x^2)=(t,x)$ and introduce the null coordinates $u=t-x$ and $v=t+x$, for which the metric on the orbit space takes the form
$g_{\mu\nu} dx^\mu dx^\nu=-f du dv$. Then the local transformations of these coordinates $u\to p(u)$, $v\to q(v)$, where $p(u)$ and $q(v)$ are arbitrary real functions  with $p^\prime>0$ and $q^\prime>0$, leaves this form of metric unchanged.
The corresponding new time-like and space-like coordinates are
$\widetilde{x}^1=\frac12(p(u)+q(v))$ and $\widetilde{x}^2=\frac12(-p(u)+q(v))$ respectively. On the other hand, the function $w(x^1,x^2)$ satifies the linear equation (\ref{wequation}). In the null coordinates $u$ and $v$ the linear equation for $w$ takes the form $w_{uv}=0$ and therefore, its solutions possess the structures $w=\widetilde{p}(u)+\widetilde{q}(v)$ with some functions $\widetilde{p}(u)$ and $\widetilde{q}(v)$. If we choose in the mentioned just above coordinate transformation $p(u)=\pm \widetilde{p}(u)$ and $q(v)=\pm \widetilde{q}(v)$, where the sign is chosen appropriately to have $p^\prime>0$ and $q^\prime>0$, we obtain that in new coordinates $(\widetilde{x}^1,\widetilde{x}^2)$ the function $w(x^1,x^2)$ takes one of the simplest forms $w=x^1$ or $w=x^2$ (where we omit the $\widetilde{}$ on the transformed coordinates). In these cases, we obtane  physically different types of fields -- e.g., $w=t$ may correspond to time-dependent cosmological solutions, while $w=x$ may correspond to spatially inhomogeneous static fields.

\paragraph{\rm\textit{\underline{\textit{In the elliptic case ($\epsilon=-1$),}}}} we may choose $\epsilon_1=1$ and $(x^1,x^2)=(x,y)$ and introduce two complex conjugated to each other coordinates $\xi=x+i y$ and $\eta=x-i y$, for which the metric on the orbit space takes the form
$g_{\mu\nu} dx^\mu dx^\nu=f d\xi d\eta$. Then the local transformations of these coordinates $\xi\to p(\xi)$, $\eta\to q(\eta)$, where $\overline{p}(\overline{\xi})$ and $q(\eta)$ are arbitrary holomorphic functions, leaves this form of metric unchanged. The corresponding new space-like coordinates are
$\widetilde{x}^1=\frac12(p(\xi)+q(\eta))$ and $\widetilde{x}^2=\frac1{2 i}(-p(\xi)+q(\eta))$. On the other hand, $w(x^1,x^2)$ satisfies the linear equation (\ref{wequation}). In the coordinates $\xi$ and $\eta$ the linear equation for $w$ takes the form $w_{\xi\eta}=0$ and therefore, its solutions is $w=\widetilde{p}(\xi)+\widetilde{q}(\eta)$ with some functions $\widetilde{p}(\xi)$ and its complex conjugate $\widetilde{q}(\eta)$. After the coordinate transformation with $p(\xi)=\widetilde{p}(\xi)$ and $q(\eta)=\widetilde{q}(\eta)$, the function $w(x^1,x^2)$ takes one of the simplest forms $w=x^1$ or $w=x^2$ (as before, we also omit $\widetilde{}$ on the transformed coordinates). From physical point of view, this may correspond to static fields with plane, or cylindrical, or, may be, some other spatial  symmetry.

\subsubsection{Simplest examples of solutions for classes of fields (I), (II), (III) }
To construct really simplest examples of solutions for these classes of fields, we use in the expressions derived previously one more ansatz $k_o=0$, that leads immediately to the following very strong restrictions
\[k_o=0\qquad\Rightarrow\qquad\psi=0\qquad\Rightarrow\qquad \mathcal{K}_1=\mathcal{K}_2=0.
\]
In this case, the equation (\ref{alphaeqnkon}) admits a solution with an arbitrary constant $c_o$
\begin{equation}\label{alphasolution}
\alpha=c_o w^{-2/n},
\end{equation}
where, as before, the parameter $n$ for different cases takes the values  $n=1$ for the class (I), $n=-6$ for the class (II) and $n=-3$ for the class (III).
The solutions of the Einstein and Einstein - Maxwell field equations corresponding to the choice (\ref{alphasolution}) can be obtained from the solution (\ref{solutionI-II-III}) -- (\ref{Phisolution}), if we put there  $k_o=0$ and take the limit $R_o\to 0$. Besides that, to obtain a correct (Lorentzian) signature of the metric, it is necessary to make an appropriate choice of the signs $\epsilon$, $\epsilon_0$ and $\epsilon_1$, as well as of the simplest form of $w$ such as $w=x^1$ or $w=x^2$ (in the metrics given below, we denote the coordinates $(x^1,x^2,x^3,x^4)$ respectively as $(t,x,y,z)$).

\paragraph{\rm\textit{\underline{\textit{Class (I): Vacuum metrics with a cosmological constant.}}}} In this case, for $\Lambda > 0$ we choose $\epsilon=1$, $\epsilon_0=1$, $\epsilon_1=-1$, $w=x^1=t$ and, after some rescalings, we obtain
\[ds^2=t^{-2}\bigl[\dfrac{3}{\Lambda}(-dt^2+dx^2)+dy^2+dz^2\bigr],
\]
and for $\Lambda < 0$ we choose $\epsilon=1$, $\epsilon_0=1$, $\epsilon_1=-1$, $w=x^2=x$, that leads to the metric
\[ds^2=x^{-2}\bigl[\dfrac{3}{(-\Lambda)}(-dt^2+dx^2)+dy^2+dz^2\bigr].
\]
It is easy to see that these metrics represent respectively de-Sitter and anti-de Sitter metrics which cover certain parts of these space-times.

\paragraph{\rm\textit{\underline{\textit{Class (II): Vacuum metrics with non-orthogonally-transitive isometry groups $\mathcal{G}_2$.}}}}\hfill\\
In the simplest case of metrics, for which the orbits of the isometry group  $\mathcal{G}_2$ do not admit the existence of 2-surfaces orthogonal to the orbits\footnote{The isometry groups $\mathcal{G}_2$, which orbits admit the existence of 2-surfaces orthogonal to the orbits are called sometimes as orthogonally-transitive ones.},
we can choose $\epsilon=1$, $\epsilon_0=1$, $\epsilon_1=-1$, $w=x^2=x$ Ё $\omega^3{}_1=\omega^{}_1=\omega^4{}_2=0$. the corresponding solution is
\[
ds^2=s_o^2 x^{-2/3}(-dt^2+dx^2)+x^{2/3}(dy+s_o x^{-2/3} dt)^2+dz^2,
\]
where $s_o$ is an arbitrary constant. After an obvious coordinate transformation $\{t,x,y,z\}\to\{T,X,Y,Z\}$, this metric can be reduced to  the following form
\[
ds^2= 2(dT+X dY) dY+dX^2+dZ^2.
\]
However, rather simple calculations show that for this metric, as well as for all metrics of the class (II), for which $k_o=0$, the Riemann tensor vanishes and therefore, these metrics are flat. This means that we obtained here the examples of non-orthogonally-transitive subgroups of the isometry group of the Minkowski space-time. A bit more complicate calculations for the metrics of the class (II) with  $k_o\ne 0$, described by the expressions  (\ref{parameters})--(\ref{omegasolution}) with $n=-6$, show that these metrics are vacuum (i.e. the Ricci tensor for these metrics vanishes), but the corresponding Riemann tensor does not vanishes and therefore, we obtain in this case the solutions of vacuum Einstein equations from the class (II) in curved space-times.

\paragraph{\rm\textit{\underline{\textit{Class (III): Solution with electromagnetic non-dynamical degrees of freedom.}}}}\hfill\\
Choosing $\epsilon=1$, $\epsilon_0=1$, $\epsilon_1=-1$, $w=x^2=x$ and $\Phi_2=0$ we obtain the solution
\[\left\{\begin{array}{l}
ds^2=q_o \overline{q}_o x^{-2/3}(-dt^2+dx^2)+x^{2/3}(dy^2+dz^2)\\[1ex]
\Phi_t=i q_o x^{-1/3}.
\end{array}\right.
\]
An interesting property of this solution is that in this space-time, in contrast to electrovacuum fields described by integrable reductions of Einstein - Maxwell equations, the electric and magnetic fields possess the components orthogonal to the orbits of the isometry group $\mathcal{G}_2$.

Thus, even the simplest examples of vacuum and electrovacuum solutions given above show that the presence of non-dynamical degrees of freedom of gravitational and electromagnetic fields can give rise to some non-trivial solutions which may be interesting from physical as well as from geometrical points of view.

\subsection{Non-dynamical degrees of  freedom and integrability}
\medskip
Here we show that vanishing of non-dynamical degrees of freedom of gravitational and electromagnetic fields is a sufficient condition for  Einstein - Maxwell equations for electrova\-cuum space-times with Abelian isometry group $\mathcal{G}_2$ to reduce to completely integrable system.

In the previous sections, the non-dynamical degrees of freedom of gravitational and electromagnetic fields were identified respectively with the metric functions $\omega^a{}_\mu$ and the components $\Phi_\mu$ of a complex electromagnetic vector potential and characterised by a set of constant parameters which consists of the components of a constant two-dimensional vector $\ell_a$, a complex constant $e_o$ and a cosmological constant $\Lambda$.\footnote{It is wondering that the "deformation" of the structure of $\mathcal{G}_2$-symmetry reduced Einstein - Maxwell equations due to the presence of non-dynamical degrees of freedom of fields is very similar to that caused by the presence of a cosmological constant. This allows us to consider (formally, at least) the cosmological constant as one more  non-dynamical degree of freedom of gravitational field.} In general, these parameters enter the symmetry reduced dynamical equations and make the structure of these equations rather complicate. However, for a special choice of the values of these constant parameters
\begin{equation}\label{zeroparams}
\Lambda=0,\quad \ell_a=0,\quad e_o=0,
\end{equation}
all the mentioned above non-dynamical degrees of freedom become pure gauge and vanish after appropriate coordinate and gauge transformations. Indeed,
as one can see from the equations (\ref{Phimu}) and (\ref{omegas}), the choice of constants (\ref{zeroparams}) leads to he equations
\[\varepsilon^{\mu\nu}\partial_\mu\omega^a{}_\nu=0,\qquad
\varepsilon^{\mu\nu}\partial_\mu\Phi_\nu= \varepsilon^{\mu\nu}\partial_\mu\Phi_c\,\omega^c{}_\nu.
\]
The first of these equations means that locally the functions $\omega^a{}_\mu$  possess the structure $\omega^a{}_\mu=\partial_\mu\omega^a$ and therefore, there exists a coordinate transformation $x^a\to x^a-\omega^a(x^\gamma)$ which leads to the condition $\omega^a{}_\mu=0$. Then the second of the equations given just above implies $\varepsilon^{\mu\nu}\partial_\mu\Phi_\nu=0$ and therefore, the functions $\Phi_\mu$ possess locally the structure $\Phi_\mu=\partial_\mu \phi(x^\gamma)$ and after the gauge transformation of electromagnetic field potential $\Phi_\mu\to\Phi_\mu-\partial_\mu \phi$ the corresponding $\Phi_\mu$ vanish.
Thus, for the parameters (\ref{zeroparams}) all non-dynamical degrees of freedom vanish,
\begin{equation}\label{Intcond}
\omega^a{}_\mu=0,\qquad  \Phi_\mu=0,
\end{equation}
and the space-time metric and complex vector electromagnetic potential  (\ref{gFComponents}) take the forms
\begin{equation}\label{BlockDiagonal}
\left.\begin{array}{l}
ds^2=f\, \eta_{\mu\nu} dx^\mu dx^\nu +g_{ab} dx^a dx^b,\\[1ex]
\Phi_i=\{0,\,0,\,\Phi_a\},
\end{array}\qquad\right\Vert\qquad
\begin{array}{l}
\mu,\nu,\ldots=1,2\\[0.5ex]
a,b,\ldots=3,4
{}
\end{array}
\end{equation}
where $f$, $g_{ab}$ and $\Phi_a$ depend on $x^\mu=\{x^1,\, x^2\}$ only.

This class of metrics and electromagnetic fields do not describe some physical and geometrical features of gravitational and electromagnetic fields which can take place within a general class of electrovacuum space-times which admit the Abelian isometry group $\mathcal{G}_2$. These restrictions include the  grevitational fields with non-orthoginally-transitive groups of isometries $\mathcal{G}_2$ and some more complicate structures of electromagnetif fields.\footnote{Such structure of complex electromagnetic potential in the hyperbolic case, i.e. for  time-dependent fields give rise to electric and magnetic fields which directions are tangent to the orbits of  $\mathcal{G}_2$, while the electromagnetic non-dynamical degrees of freedom, if not vanish,  correspond to electric and magnetic fields with the components in the spatial direction orthogonal to the orbits of $\mathcal{G}_2$. In the elliptic case, i.e. for stationary fields with one spatial symmetry, this structures of metric and complex electromagnetic potential give rise to electric and magnetic fields with the components in the directions orthogonal to the orbits of  $\mathcal{G}_2$, while the non-dynamical electromagnetic degrees of freedom, if not vanish, give rise to  electric and magnetic fields with the components in the directions tangent to the orbits of  $\mathcal{G}_2$.}  However, the well known remarkable property of this class of fields is that the Einstein - Maxwell equations in this case simplify considerably and become completely integrable. Just this class of electrovacuum fields and integrability of  the corresponding Einstein - Maxwell equations  were used by different authors in numerous studies of recent decades of many aspects of behaviour of strong gravitational  and electromagnetic fields and their nonlinear interactions.

\section{Concluding remarks}

In this paper a structure of field equations for the general classes of vacuum and electrovacuum fields for which the space-time admits two-dimensional Abelian isometry group and the electromagnetic field possess the same symmetry. These classes of fields include all fields described by the known integrable reductions of vacuum Einstein equations and electrovacuum Einstein - Maxwell equations. However, the fields in these classes can include also the components, which are called here as the non-dynamical degrees of freedom of gravitational and electro\-mag\-netic fields, and which are characterised by a set of constants arising as the constants of integration of the constraint equations. These constants  in general enter also the dynamical equations for gravitational and electromagnetic fields, change essentially these equations and destroy their known integrable structure.  In the paper, we describe different forms of $\mathcal{G}_2$-symmetry reduced dynamical equations modified appropriately for the case of presence of non-dynamical degrees of frredom. These are the modified Ernst equations for complex scalar potentials and matrix sel-dual Kinnersley equatyions. The simplest examples are presented for solutions with different non-vanishing non-dynamical degrees of freedom.

It is worth mentioning here once more a surprising analogy between all considered above cases of a presence of non-dynamical degrees of freedom of gravitational and electromagnetic fields (from one side) and the case of presence of a cosmological constant (from the other side) which change
the structures of $\mathcal{G}_2$-symmetry reduced Einstein - Maxwell equations (in comparison with the integrable reductions of these equations) in a very similar ways which do not allow to generalize for these cases the well known methods used for solution of integrable reductions of these equations arising in the absence of the mentioned above factors. The simplest examples of solutions with non-dynamical degrees of freedom of fields were given.

It was shown also that for a special choice of values of the constant parameters which characterise non-dynamical degrees of freedom, these degrees of freedom occur to be pure gauge and these can vanish for appropriate choice of coordinates and gauge transformations of fields. Thus, it occurs that in the case of existence of two-dimensional Abelian group of isometries, the vanishing of all non-dynamical degrees of freedom of electrovacuum fields provide the sufficient condition for integrability of this symmetry reduced vacuum Einstein equations and electrovacuum Einstein - Maxwell equations.‚
We note also that the similar non-dynamical degrees of freedom of fields exist also in the other cases of Einstein's field equations which admit the integrable two-dimensional reductions. In this cases, the similar assumptions about the space-time symmetry and vanishing of all non-dynamical degrees of freedom also occur to be the sufficient conditions for integrability of the corresponding symmetry reduced dynamical equations. In particular, these are e.g., the Einstein - Maxwell - Weyl equations which describe the nonlinear interaction of gravitational, electromagnetic and massless two-component Weyl spinor field \cite{Alekseev:1983}, as well as two-dimensional reductions of Einstein equations for some string gravity models in space-times of four or higher dimensions \cite{Alekseev:2013}.

It is interesting to note also that for all known integrable reductions of Einstein's field equations the condition of "harmonical" structure is satisfied for the function $\alpha(x^1,x^2)$, which determines the element of area on the orbits of the isometry group $\mathcal{G}_2$ (in four-dimensional space-times) or $\mathcal{G}_{D-2}$ (for gravity models in $D$-dimensional space-times with $D > 4$), while in the other (more general) cases which were considered in the present paper and which (presumably) are not integrable with the known methods, the function $\alpha$ is a dynamical variable which should satisfy a complicate nonlinear equations. Therefore, if for some two-dimensional reduction of Einstein's field equations the function $\alpha$ is a "harmonic" function, this can be an indication of possible complete integrability of these reduced dynamical equations for the fields with vanishing of all non-dynamical degrees of freedom.

\subsection*{Acknowledgments}
This work is supported by the Russian Science Foundation under grant 14-50-00005.

\end{document}